%% file: exp-ml-mkic-d4.tex
\documentclass[a4paper,UKenglish,autoref,11pt]{article}

\usepackage{microtype}

\makeatletter
\renewcommand{\maketitle}{\bgroup\setlength{\parindent}{0pt}
\begin{flushleft}
  \textbf{\LARGE\@title}
  
  \vspace{.5cm}
  \@author
\end{flushleft}\egroup
}
\makeatother

\usepackage{soul}

\title{New Exponential Size Lower Bounds against Depth Four Circuits of Bounded Individual Degree}

\author{\textbf{Suryajith Chillara},\\CRI, University of Haifa, Israel.\\\texttt{\small suryajith@cmi.ac.in}}

\date{}

\usepackage[dvipsnames,svgnames,x11names,hyperref]{xcolor}
\usepackage[framemethod=TikZ]{mdframed}
\usepackage{tikz,epigraph,verbatim,thmtools,thm-restate,ccicons,setspace,microtype}

\usepackage{amsmath}
\usepackage{amssymb}
\usepackage{amsthm}
\usepackage{fullpage}

\usepackage{todonotes}
\usepackage[bitstream-charter]{mathdesign}

\usepackage{eulervm}
\usepackage[T1]{fontenc}

\usepackage[colorlinks]{hyperref}
\hypersetup{
  linkcolor=[rgb]{0.3,0.3,0.6},
  citecolor=[rgb]{0.2, 0.6, 0.2},
  urlcolor=[rgb]{0.6, 0.2, 0.2}
}

\usepackage[capitalise]{cleveref}
\allowdisplaybreaks

\newtheorem{theorem}{Theorem}

\newtheorem{lemma}[theorem]{Lemma}
\newtheorem{observation}[theorem]{Observation}
\newtheorem{proposition}[theorem]{Proposition}
\newtheorem{definition}[theorem]{Definition}
\newtheorem{claim}[theorem]{Claim}

\newcommand{\mrc}{multi-$\delta$-ic\ }
\newcommand{\mrcvec}{multi-$\overline{\delta}$-ic\ }

\newcommand{\supp}[1]{\operatorname{\mathrm{Supp}}(#1)}

\newcommand{\monsuppsize}[1]{\abs{\operatorname{\mathrm{MonSupp}}(#1)}}
\newcommand{\avg}[2]{\operatorname{\mathbb{E}}_{#1}\insquare{#2}}
\newcommand{\dist}{\operatorname{\mathrm{dist}}}
\newcommand{\mult}{\operatorname{\mathrm{mult}}}
\newcommand{\ml}{\mathrm{ML}}
\newcommand{\nml}{\mathrm{NonML}}

\newcommand{\M}{\mathcal{M}}
\newcommand{\CM}[1]{\mathrm{PSSPD}_{#1}}
\newcommand{\tiln}{ n }

\newcommand{\bldiff}{\operatorname{\mathrm{BlockDiff}}}
\input{minimacros}

\makeatletter

\begin{document}
\maketitle

\vskip1mm
\hrule
\vskip1mm
\begin{abstract}
  Kayal, Saha and Tavenas [Theory of Computing, 2018] showed that for all large enough integers $n$ and $d$ such that $d\geq \omega(\log{n})$, any \emph{syntactic} depth four circuit of bounded individual degree $\delta = o(d)$ that computes the Iterated Matrix Multiplication polynomial ($\IMM_{n,d}$) must have size $n^{\Omega\inparen{\sqrt{d/\delta}}}$. Unfortunately, this bound deteriorates as the value of $\delta$ increases. Further, the bound is superpolynomial only when $\delta$ is $o(d)$. It is natural to ask if the dependence on $\delta$ in the bound could be weakened. Towards this, in an earlier result [STACS, 2020], we showed that for all large enough integers $n$ and $d$ such that $d = \Theta(\log^2{n})$, any \emph{syntactic} depth four circuit of bounded individual degree $\delta\leq n^{c}$ (for a small constant $c$) that computes $\IMM_{n,d}$ must have size $n^{\Omega(\log{n})}$.

  In this paper, we make further progress by proving that for all large enough integers $n$ and $d$, and absolute constants $a$ and $b$ such that $\omega(\log^2n)\leq d\leq n^{a}$, any \emph{syntactic} depth four circuit of bounded individual degree $\delta\leq n^{b}$ that computes $\IMM_{n,d}$ must have size $n^{\Omega(\sqrt{d})}$. Our bound is obtained by carefully adapting the proof of Kumar and Saraf [SIAM J. Computing, 2017] to the complexity measure introduced in our earlier work [STACS, 2020].
	
\end{abstract}

\vskip1mm
\hrule
\vskip1mm
\section{Introduction}

Arithmetic circuits are directed acyclic graphs such that the leaf nodes are labeled by variables or constants from the underlying field, and every non-leaf node is labeled either by a $+$ or $\times$. Every node computes a polynomial by  operating on its inputs with the operand given by its label. The flow of computation flows from the leaf to the output node. We refer the readers to the standard resources~\cite{sy, github} for more information on arithmetic formulas and arithmetic circuits.

One of the central questions in Algebraic Complexity Theory is to show super polynomial size lower bounds against \emph{hard} polynomials. It is interesting to note that most polynomials of interest are multilinear. However, in the last few decades, we could not make substantial progress towards proving lower bounds for multilinear polynomials against general arithmetic circuits (cf.~\cite{sy, github}). It then is a natural strategy to prove lower bounds against restricted class of circuits. To begin with, we can restrict the polynomial computed at every gate to be multilinear. Under such a restriction, researchers made a lot of progress~\cite{nw1997, Raz, Raznc2nc1, ry08, ry09, hy, RSY08, AKV, CLS, CELS, CLS2}. Backed by this progress, we can now try to relax this restriction to circuits where the polynomial computed at every gate is of bounded individual degree.

Kayal and Saha~\cite{KS17} first studied \mrc circuits of depth three and proved exponential lower bounds. Kayal, Saha and Tavenas~\cite{KST} have extended this and proved exponential lower bounds at depth three and depth four, and superpolynomial lower bounds for homogeneous formulas. These circuits (formulas) that were considered were syntactically \mrc. That is, at any product node, any variable appears in the support of at most $\delta$ many operands, and the total of the individual formal degrees is also at most $\delta$. Henceforth, all the \mrc depth four circuits that we talk about shall be syntactically \mrc.

Recently, Kumar, Oliviera and Saptharishi~\cite{KOS} showed that proving lower bounds of the order $\exp(O(\sqrt{\delta\cdot N\log N}))$ for $N$ variate polynomials against depth four \mrc circuits is sufficient to show superpolynomial lower bounds against general \mrc circuits. This provides us further motivation to study \mrc circuits of depth four. 

Raz and Yehudayoff~\cite{ry09} showed a lower bound of $\exp(\Omega\inparen{\sqrt{d\log d}})$ against multilinear depth four circuits which compute a multilinear polynomial over $N$ variables and degree $d\ll N$ (cf.\ \cite[Footnote 9]{KST}). 
Kayal, Saha and Tavenas \cite{KST} have shown a lower bound of $\inparen{\frac{n}{\delta^{1.1}}}^{\Omega\inparen{\sqrt{\frac{d}{\delta}}}}$ for Iterated Matrix Multiplication polynomial over $d$ many $n\times n$ matrices (denoted $\IMM_{n,d}$), that is computed by a \mrc depth four circuit. This lower bound deteriorates with the increasing value of $\delta$ and is superpolynomial only when $\delta$ is $o(d)$ and is strictly less than $n^{1.1}$. This raises a natural question if the dependence on the individual degree could be improved upon. Towards this, in the same paper, they showed a lower bound of $2^{\Omega\inparen{\sqrt{N}}}$ for a $N$-variate polynomial that is not multilinear. Hegde and Saha~\cite{HS} proved a bound of $2^{\Omega\inparen{\sqrt{\frac{N\log N}{\delta}}}}$ against a $N$-variate multilinear design polynomial which is in $\VNP$, for a wider range of $\delta$ than \cite{KST}.

In an earlier result~\cite{ChiSTACS20}, we showed that for all $\delta<n^{c}$ (for a small constant $c$) and for a specific regime of degree $d=\Theta(\log^2n)$, any \mrc depth four circuit computing $\IMM_{n,d}$ must have size $n^{\Omega\inparen{\log n}}$. Even though the bound holds for a very small range of degrees and degrees much smaller, this quasipolynomial bound holds true even for individual degree much larger than the syntactic degree of the polynomial.

In this paper, we extend \cite{ChiSTACS20} and show that for all large enough integers $n$ and $d$, and absolute constants $a$ and $b$ such that $b>a$ and $\omega(\log^2n)\leq d\leq n^{a}$, any \emph{syntactic} depth four circuit of bounded individual degree $\delta\leq n^{b}$ that computes $\IMM_{n,d}$ must have size $n^{\Omega(\sqrt{d})}$.

To prove our result, we use the dimension of Projected Shifted Skew Partial Derivatives as our complexity measure. This was introduced in \cite{ChiSTACS20} and is an extension of the dimension of Shifted Skew Partial Derivatives, introduced by Kayal et al.~\cite{KST}. This extension was inspired by the definition of Projected Shifted Partial Derivatives measure of Kayal, Limaye, Saha and Srinivasan~\cite{KLSS}. The polynomial of interest is multilinear and it makes sense to project down the space of Shifted Skew Partial Derivatives to just those Shifted Skew Partial Derivatives which are multilinear.

For a polynomial $P$ defined over the variable set $Y\sqcup Z$, the dimension of Projected Shifted Skew Partial Derivatives with respect to the parameters $r'$ and $m$ can informally be defined as follows.
\begin{align*}
  \dim\inparen{\F\text{-span}\inbrace{\mult\inparen{\vecz^{= m}\cdot \sigma_Y\inparen{\partial^{=r'}_Y P}}}}
\end{align*}
Here, $\vecz^{= m}$ denotes the set of monomials over $Z$-variables of degree $m$, $\sigma_Y(f)$ sets all the $Y$ variable appearances in $f$ to $0$ and $\mult(g)$ sets the coefficients of all the nonmultilinear monomials in $g$ to $0$. Using linear algebra and setting an order on the variables, counting the dimension of the aforementioned quantity can be reduced to counting leading monomials in the space $\inparen{\F\text{-span}\inbrace{\mult\inparen{\vecz^{= m}\cdot \sigma_Y\inparen{\partial^{=r'}_Y P}}}}$.

\paragraph*{Comparison to \cite{KST, HS}:} Even though Kayal et al.\cite{KST} and us in this paper, have proved lower bounds for $\IMM_{n,d}$, when compared to Kayal et al.~\cite{KST} our bound holds for a restricted range of $d$. In \cite{KST}, they proved a lower bound for all $d\geq\log^2{n}$. In contrast, our bound holds only for degrees smaller than $n^{a}$ for an absolute constant $a$. On the other hand, for all $\omega(\log^2n) \leq d \leq n^a$, we show better quantitative bounds than \cite{KST}. In this regime of $d$, we show a bound that does not deteriorate with the increase in individual degree of the depth four circuit. Further our bound holds for a range of $\delta$ that is greater than $d$ as well, which was not the case in \cite{KST}. 

Hegde and Saha~\cite{HS} proved an exponential lower bound for a $N$-variate polynomial of degree $\Theta(N)$ which is in $\VNP$. Our result is mostly incomparable against this. 

\paragraph*{Comparison to \cite{ChiSTACS20}:} In \cite{ChiSTACS20}, we defined a polynomial $Q_n$ which is a $p$-projection of $\IMM_{n,d}$ where $d = \Theta(\log^2{n})$. We then used random restrictions $\rho$ on the variable set and reduced $Q_n$ to a polynomial $P_\rho$. At the same time we showed that with a high probability, the depth four \mrc circuit under random restrictions would reduce to a depth four \mrc circuit such that every monomial at the bottom product gate is of low support. Then using the dimension of Projected Shifted Skew Partial Derivatives as our complexity measure, we showed that any \mrc depth circuit of low bottom support that computes $P_\rho$ must have large size. Through union bound, we inferred that any \mrc depth four circuit computing $Q_n$ must be large. Since $Q_n$ was a $p$-projection of $\IMM_{n,\Theta(\log^2n)}$, we get that any \mrc depth four circuit computing $\IMM_{n,\Theta(\log^2n)}$ must be large.

An important point to note in \cite{ChiSTACS20}, all the Skew Partial Derivatives were multilinear monomials. Thus, a leading monomial is given by the projected shift of the only multilinear monomial. This analysis fell short of giving bounds that were anything better than quasipolynomial. To overcome this hurdle, in this paper we ensure that the Skew Partial Derivatives are in fact multilinear polynomials over a large set of monomials. This puts us in a similar situation as in \cite{KS14}. The key observation in \cite{KS14} (cf. \cite[Section 20.3]{github}) was that if $\beta$ is a leading monomial in $\sigma_Y(\partial_\alpha(P))$, then for some $\gamma$, $\gamma\cdot\beta$ need not be a leading monomial in $\mult(\gamma\cdot\sigma_Y(\partial_\alpha(P)))$.

\paragraph*{Similarity to \cite{KS14}:} As discussed above, we extend the random restrictions and the careful analysis of Kumar and Saraf~\cite{KS14} and adapt it to our complexity measure. We work with a different set of parameters than \cite{KS14} and thus, we build on their already \emph{careful} analysis and refine it in certain places to suit our needs. 

In this paper, we made a conscious decision to use a notation which is as close to that of \cite{KS14} as possible to help those readers who are already acquainted with the proof of \cite{KS14}.

\subsection*{Proof overview:}

The proof strategy is very similar to the previous work~\cite{FLMS, KS14, KST, HS, ChiSTACS20}. We first pick a random restriction $V$ of the variables from a carefully crafted distribution $D$. We then show that with a high probability, a \mrc depth four circuit under such a restriction reduces to a \mrc depth four circuit of low bottom $Z$-support. Let $\Phi|_V$ be the circuit obtained after the restriction $V\leftarrow D$.  
Let $T_1, T_2, \dots, T_s$ be the terms corresponding to the product gates feeding into the output sum gate of $\Phi|_V$. The output polynomial is obtained by adding the terms $T_1, T_2, \dots, T_s$. Note that each of these $T_i$'s is a product of low $Z$-support polynomials $Q_{i,j}$, that is, every monomial in these $Q_{i,j}$'s is supported on a small set of $Z$ variables (say $t$ many). One major observation at this point is to see that there can be at most $\abs{Z}\cdot r$ many factors in any of the $T_i$'s with non-zero $Z$-support.

From \cite{ChiSTACS20}, we get that the dimension of Projected Shifted Skew Partial Derivatives for \emph{small} \mrc depth four circuits of low bottom $Z$-support is \emph{not too large}. We then show that the Projected Skew Partial Derivative space of a polynomial $\IMM_{n,d}|_V$ is \emph{large} using a Leading Monomial approach (cf. \cite[Section 20.3]{github}). Thus, we infer that if $\Phi|_V$ were to compute $\IMM_{n,d}|_V$, then $\Phi|_V$ cannot be \emph{small}. Then we lift this argument to show that if $\Phi$ were to compute $\IMM_{n,d}$, then $\Phi$ cannot be \emph{small}.

\section{Preliminaries}
\paragraph*{Notation:}
\begin{itemize}
\item For a polynomial $f$, we use $\partial^{=r'}_Y(f)$ to refer to the space of partial derivatives of order $r'$ of $f$ with respect to monomials of degree $r'$ in $Y$.
\item We use $\vecz^{=m}$ and $\vecz^{\leq m}$ to refer to the set of all the monomials of degree equal to $m$ and at most $m$, respectively, in $Z$ variables. 
\item We use $\vecz_{\ml}^{\leq m}$ to refer to the set of all the multilinear monomials of degree at most $m$ in $Z$ variables.
\item We use $\vecz_{\nml}^{\leq m}$ to refer to the set of all the non-multilinear monomials of degree at most $m$ in $Z$ variables.
\item For a monomial $m$ we use $\monsuppsize{m}$ to refer to the size of the set of variables that appear in it.
\item For a polynomial 	$f$, we use $\monsuppsize{f}$ to refer to the maximum $\monsuppsize{m}$ over all monomials in $f$.
\end{itemize}

\paragraph*{Depth four circuits:} A depth four circuit (denoted by $\SPSP$) over a field $\F$ and variables $\inbrace{x_1, x_2, \dots, x_n}$ computes polynomials which can be expressed in the form of sums of products of polynomials. That is, $ \displaystyle\sum_{i=1}^s\prod_{j=1}^{d_i}Q_{i,j}(x_1,\dots,x_n)$
for some $d_i$'s. A depth four circuit is said to have a bottom support of $t$ (denoted by $\SPSPsupp{t}$) if it is a depth four circuit and all the monomials in each polynomial $Q_{i,j}$ are supported on at most $t$ variables.

\paragraph*{Multi-$\delta$-ic arithmetic circuits:} 
  Let $\overline{\delta}=(\delta_{1}, \delta_{2}, \cdots, \delta_{N})$. An arithmetic circuit $\Phi$ is said to be a syntactically \mrcvec circuit if for all product gates $u\in \Phi$ and $u = u_{1}\times u_{2}\times \cdots \times u_{t}$,  each variable $x_i$ can appear in at most $\delta_i$ many of the $u_i$'s ($i\in[t]$). Further the total formal degree with respect to every variable $x_i$ over all the polynomials computed at $u_1, u_2, \cdots, u_t$, is bounded by $\delta_i$, i.e.\ $\sum_{j\in[t]}\deg_{x_{i}}(f_{u_{j}}) \leq \delta_i$ for all $i\in[N]$. If $\overline{\delta} = (\delta, \delta, \cdots, \delta)$, then we simply refer to them as \mrc circuits.

\paragraph*{Complexity Measure:} 

Let the variable set $X$ be partitioned into two fixed, disjoint sets $Y$ and $Z$ such that $\abs{Y}$ is a magnitude larger than $\abs{Z}$. Let $\sigma_Y: \F[Y\sqcup Z] \mapsto \F[Z]$ be a map such that for any polynomial $f(Y,Z)$, $\sigma_Y(f)\in \F[Z]$ is obtained by setting every variable in $Y$ to zero by it and leaving $Z$ variables untouched. Let $\text{mult}: \F[Z]\mapsto \F[Z]$ be a map such that for any polynomial $f(Y,Z)$, $\text{mult}(f)\in \F[Z]$ is obtained by setting the coefficients of all the non-multilinear monomials in $f$ to 0 and leaving the rest untouched. We use $\vecz^{= m}\cdot \sigma_Y(\partial_Y^{=r'}f)$ to refer to the linear span of polynomials obtained by multiplying each polynomial in $\sigma_Y(\partial_Y^{=r'}f)$ with monomials of degree $m$ in $Z$ variables. We will now define our complexity measure, Dimension of Projected Shifted Skew Partial Derivatives (denoted by $\CM{r',m}$) as follows.
\begin{align*}
  \CM{r',m}(f) = \dim\inparen{\F\text{-span}\inbrace{\mult\inparen{\vecz^{= m}\cdot \sigma_Y\inparen{\partial^{=r'}_Y f}}}}
\end{align*}

This is a natural generalization of Shifted Skew Partial Derivatives of \cite{KST}. The following proposition is easy to verify.

\begin{proposition}[Sub-additivity]\label{prop:subadd}
	Let $r'$ and $m$ be integers. Let the polynomials $f, f_1, f_2$ be such that $f = f_1 + f_2$. Then, $\CM{r', m}(f) \leq \CM{r', m}(f_1)+\CM{r', m}(f_2)\,.$
\end{proposition}

\paragraph*{Monomial Distance:} We recall the following definition of distance between monomials from \cite{CM19}.
\begin{definition}[Definition 2.7, \cite{CM19}]
  Let $m_1, m_2$ be two monomials over a set of variables. Let $S_1$ and $S_2$ be the multisets of variables corresponding to the monomials $m_1$ and $m_2$ respectively. The distance $\dist(m_1, m_2)$ between the monomials $m_1$ and $m_2$ is the $\min\{|S_1|-|S_1\cap S_2|, |S_2|-|S_1\cap S_2|\}$ where the cardinalities are the order of the multisets. 
\end{definition}

For example, let $m_1 = x_1^2x_2x_3^2x_4$ and $m_2 = x_1x_2^2x_3x_5x_6$. Then $S_1 = \{x_1, x_1, x_2, x_3, x_3, x_4\}$, $S_2 = \{x_1, x_2, x_2, x_3, x_5, x_6\}$, $|S_1|=6$, $|S_2|=6$ and $\dist(m_1, m_2) = 3$. It is important to note that two distinct monomials could have distance $0$ between them if one of them is a multiple of the other and hence the triangle inequality does not hold.

\paragraph*{Polynomial Family:}
Let $X^{(1)}, X^{(2)}, \ldots, X^{(d)}$ be $d$ generic $n\times n$ matrices defined over disjoint set of variables. For any $k\in[d]$, let $x_{i,j}^{(k)}$ be the variable in the matrix $X^{(k)}$ indexed by $(i,j)\in[n]\times[n]$. The Iterated Matrix Multiplication polynomial, denoted by the family $\{\IMM_{n,d}\}$, is defined as follows.  
  \begin{align*}
    \IMM_{n,d}(X) = \sum_{i_1, i_2, \ldots, i_{d-1} \in [n]}x_{1,i_1}^{(1)}x_{i_1,i_2}^{(2)}\dots x_{i_{(d-2)},i_{(d-1)}}^{(d-1)}x_{i_{(d-1)},1}^{(d)}.
  \end{align*}

The following lemma (from \cite{GKKS}) is key to the asymptotic estimates required for the lower bound analyses. 
\begin{lemma}[Lemma 6, \cite{GKKS}]
  \label{lem:bin-gkks}
  Let $a(n), f(n), g(n) : \mathbb{Z}_{\geq 0} \rightarrow \mathbb{Z}_{\geq 0}$ be integer valued functions such that $(f+g) = o(a)$. Then, 
  \begin{align*}
    \ln \frac{(a+f)!}{(a-g)!} = (f+g)\ln a \pm O\left(\frac{(f+g)^2}{a}\right)
  \end{align*}
\end{lemma}

As in \cite{KS14}, we also use the following strengthening of the Principle of Inclusion and Exclusion in our proof. 
\begin{lemma}[Strong Inclusion-Exclusion~\cite{KS14}]
  \label{lem:IE-KumarSaraf}
  Let $W_1, W_2, \cdots, W_t$ be subsets of a finite set $W$. For a parameter $\lambda \geq 1$, let $\sum_{\substack{i,j\in[t]\\i\neq j}}\abs{W_i\cap W_j}\leq \lambda\sum_{i\in[t]}\abs{W_i}\,.$
  Then, $\abs{\cup_{i\in[t]} W_i} \geq \frac{1}{4\lambda} \sum_{i\in[t]}\abs{W_i}\,.$

\end{lemma}

We also need the following form of generalized Hamming bound~\cite[Section 1.7]{GAS}.
\begin{lemma}\label{clm:hammingbnd}
  
  For every $\Delta_0<2r'$, there exists a subset $\mathcal{P}_{\Delta_0} \subset [n]^{2r'}$ of size $\frac{n^{2r'-\Delta_0/2}}{\frac{\Delta_0}{2}\cdot{2r'\choose {0.5\Delta_0}}}$ such that for all $\veca, \veca' \in \mathcal{P}_{\Delta_0}$,  $\dist(\veca, \veca') \geq \Delta_0$.
\end{lemma}

\subsection*{Leading Monomial Approach of \cite{KS14}}\label{sec:LeadingMonomials}
Let $P$ be any polynomial defined over the sets of variables $Y\sqcup Z$, of degree $d$. Let $\M$ be a set of suitably chosen monomials of degree equal to $r'$ over $Y$ variables. Let $\sigma_Y$ applied to a polynomial, set all its $Y$ variables to zero. For any monomial $\alpha\in\M$, let $M(\alpha)$ be the set of monomials in the support of the polynomial $\sigma_Y(\partial_\alpha(P))$. Note that all the monomials in $M(\alpha)$ are supported only on $Z$ variables. For any $\beta \in M(\alpha)$, let
\begin{align*}
  A_m(\alpha,\beta) &= \inbrace{\gamma'\mid \supp{\gamma'}=\deg(\gamma')=d-r'+m~\text{and}~\exists \gamma~\text{s.t}~\gamma'=\LM\inparen{\gamma\cdot\sigma_Y(\partial_\alpha(P))} = \gamma\cdot\beta}.
\end{align*}
\begin{proposition}[Equation 20.13, \cite{github}]
  Let $m,r'$ be integers. Then,
  \begin{align*}
    \CM{r',m}(P) &\geq \abs{\bigcup_{\substack{\alpha \in \M\\\beta\in M(\alpha)}}A_m(\alpha,\beta)}\,.
  \end{align*}
\end{proposition}
Using the Inclusion-Exclusion principle we get the following.
\begin{align*}
  \abs{\bigcup_{\substack{\alpha \in \M\\\beta\in M(\alpha)}}A_m(\alpha,\beta)} \geq \sum_{\substack{\alpha\in \M\\\beta\in M(\alpha)}}\abs{A_m(\alpha,\beta)} - \sum_{\substack{\alpha, \alpha'\in \M\\\beta\in M(\alpha)\\\gamma\in M(\alpha')\\(\alpha, \beta) \neq (\alpha', \gamma)}}\abs{A_m(\alpha,\beta)\cap A_m(\alpha',\gamma)}.
\end{align*}
For any $\beta \in M(\alpha)$, let
\begin{align*}
  S_m(\alpha, \beta) &= \inbrace{\gamma\mid \deg(\gamma) = \supp{\gamma} = m~\text{and}~\abs{\supp{\gamma}\cap\supp{\beta}}=\emptyset}.
\end{align*}
\begin{lemma}[Lemma 20.17, \cite{github}]
  For all $\alpha\in \M$, 
  \begin{align*}
    \sum_{\beta\in M(\alpha)}\abs{A_m(\alpha,\beta)} &\geq \abs{\bigcup_{\beta\in M(\alpha)}S_m(\alpha,\beta)}.
  \end{align*}
\end{lemma}
Again, by using the Inclusion-Exclusion principle, we get that
\begin{align*}
  \abs{\bigcup_{\beta\in M(\alpha)}S_m(\alpha,\beta)}
    &\geq \sum_{\beta\in M(\alpha)}\abs{S_m(\alpha,\beta)} - \sum_{\substack{\beta,\gamma\in M(\alpha)\\\beta\neq\gamma}}\abs{S_m(\alpha,\beta)\cap S_m(\alpha, \gamma)}.
\end{align*}

For a polynomial $P$ and monomials $\alpha, \alpha' \in \M$, let $T_1(P,\alpha), T_2(P,\alpha)$ and $T_3(P, \alpha, \alpha')$ be defined as follows.
\begin{align*}
  T_1(P, \alpha) &= \sum_{\beta\in M(\alpha)}\abs{S_m(\alpha,\beta)}\\
  T_2(P, \alpha) &=  \sum_{\substack{\beta,\gamma\in M(\alpha)\\\beta\neq\gamma}}\abs{S_m(\alpha,\beta)\cap S_m(\alpha, \gamma)}\\
  \text{and}\quad T_3(P, \alpha, \alpha') &= \sum_{\substack{\alpha, \alpha'\in \M\\\beta\in M(\alpha)\\\gamma\in M(\alpha')\\(\alpha, \beta) \neq (\alpha', \gamma)}}\abs{A_m(\alpha,\beta)\cap A_m(\alpha',\gamma)}.
\end{align*}
Further, let $T_1(P), T_2(P)$ and $T_3(P)$ be defined as follows.
\begin{align*}
  T_1(P) &= \sum_{\alpha \in \M}T_1(P, \alpha)\\
  T_2(P) &= \sum_{\alpha \in \M}T_2(P, \alpha)\\
 \text{and}\quad T_3(P) &= \sum_{\alpha, \alpha' \in \M}T_3(P, \alpha, \alpha').
\end{align*}
Thus we get that $\CM{r',m}(P) \geq T_1(P) - T_2(P) - T_3(P)\,.$

The crux of the paper is to efficiently bound $\CM{r',m}(P)
$ for an explicit polynomial $P$ by obtaining suitable bounds on $T_1(P)$, $T_2(P)$, and $T_3(P)$.

\subsection*{Choice of parameters:}
\begin{itemize}
\item Set $\varepsilon' = 0.340$.
\item Set $\eta = 0.05$ and thus $\varepsilon = \varepsilon' - \eta=0.29$.
\item Let $r= 2\tau r'$ and we set $\tau = 0.08$.
\item Let $c_0$ be a constant that is at most $0.016$.
\item Let us set $t$ to $2\times 10^{-5}k'$.
\item Since $(2k'+3)r' = d$, we can set $k'$ and $r'$ to be of the same order and thus are $\approx O(\sqrt{d})$. Let $2c_0r'\leq t$.
\item Let $k = d - 3r' = 2k'r'$.
\item Let $d\leq n^{0.01}$.
\item We need to ensure that $\nu(1-\tau) > 2\varepsilon$ and $1-\nu-\varepsilon'>0$. Thus, we set $\nu= 0.631$.
\item We want $n^\eta\cdot 2^{k'-\varepsilon'\log{n}-1} = \frac{2^{k'}}{2\cdot n^{\varepsilon}} = \inparen{\frac{N}{N-m}}^{k'}$. Let us set $m$ to $\frac{N}{2}(1-\Gamma)$ such that $\Gamma =\frac{\ln n}{\lambda \cdot k'}$ and the constant $\lambda$ is determined by the equation $(1+\Gamma)^{k'} = 2n^{\varepsilon}$. When $k'= \omega(\ln{n})$,
\begin{align*}
  2n^{\varepsilon} = (1+\Gamma)^{k'}\approx e^{\frac{\ln n}{\lambda \cdot k'}\cdot k'} = n^{\frac{1}{\lambda}}\,.
\end{align*}
\end{itemize}

\section{Deterministic and Random Restrictions}

Let the $d$ matrices be divided into $r'$ contiguous blocks of matrices $B_1, B_2, \dots, B_{r'}$ such that each block $B_i$ contains $2k'+3$ matrices and $d = (2k'+3)\cdot r'$. By suitable renaming, let us assume that each block $B_i$ contains the following matrices.
\begin{align*}
X^{(i,L,k'+1)}, \cdots, X^{(i,L,2)},X^{(i,L,1)}, X^{(i)},X^{(i,R,1)}, X^{(i,R,2)}, \cdots, X^{(i,R,k'+1)}.
\end{align*}

Let us first consider the following set of restrictions, first deterministic and then randomized. 

\subsubsection*{Deterministic Restrictions}
Let $V_0:X\mapsto Y_0\sqcup Z_0\sqcup \inbrace{0,1}$ be a deterministic restriction of the variables $X$ in to disjoint variable sets $Y_0$, $Z_0$, and $\zo$ as follows. For all $i\in[r']$, 
\begin{itemize}
\item The variables in matrix in $X^{(i)}$ are each set to a distinct $Y_0$ variable. Henceforth, we shall refer to this as $Y^{(i)}$ matrix.
\item The entries of the first row of matrix $X^{(i,L,k'+1)}$ are all set to $1$ and the rest of the matrix to $0$.
\item The entries of the first column of matrix $X^{(i,R,k'+1)}$ are all set to $1$ and the rest of the matrix to $0$.
\item The rest of the variables are all set to distinct $Z_0$ variables. Henceforth, for all $b\in\inbrace{L,R}$ and $j\in[k']$, we shall refer to the matrix $X^{(i,b,j)}$ as $Z^{(i,b,j)}$ matrix.
\end{itemize}

\subsubsection*{Random Restrictions}
Let $V_1:Y_0\sqcup Z_0\mapsto Y\sqcup Z\sqcup \inbrace{0,1}$ be a random restriction of the variables $Y_0\sqcup Z_0$ as follows.
\begin{itemize}
\item Matrix $Z^{(i,L,1)}$: For every column, pick $ n ^{\eta}$ distinct elements uniformly at random and keep these elements alive. Set the other entries in this matrix to zero.
\item Matrix $Z^{(i,R,1)}$: For every row, pick $ n ^{\eta}$ distinct elements uniformly at random and keep these elements alive. Set the other entries in this matrix to zero.
\item Matrices $Z^{(i,L,j)}$ for all $j\in[2, k'-\varepsilon'\log{n}]$: For every column, pick $2$ distinct elements uniformly at random and set all the other entries to zero. 
\item Matrices $Z^{(i,R,j)}$ for all $j\in[2, k'-\varepsilon'\log{n}]$: For every row, pick $2$ distinct elements uniformly at random and set all the other entries to zero.
  \item Matrices $Z^{(i,L,j)}$ for all $j>k'-\varepsilon'\log{n}$: For every column, pick $1$ element uniformly at random and set the other elements in that row to zero.
\item Matrices $Z^{(i,R,j)}$ for all $j>k'-\varepsilon'\log{n}$: For every row, pick $1$ element uniformly at random and set the other elements in that row to zero.
\end{itemize}

Let $D$ be the distribution of all the restrictions $V:X\mapsto Y\sqcup Z\sqcup\zo$ such that $V = V_1\circ V_0$ where $V_0$ and $V_1$ are deterministic and random restrictions respectively, as described above. Let $N$ be used to denote the number of $Z$ variables left after the restriction and $N = 2r'n(n^{\eta}+2(k'-\varepsilon'\log{n}-1)+\varepsilon'\log{n})$. We set the value of $k'$ to be much smaller than $n^\eta$ and thus, $N=O(n^{1+\eta}\cdot r')$.

\subsection*{Effect of restrictions on Depth Four Multi-$\delta$-ic Circuits}
\begin{lemma}\label{lem:cktrandrest}
  Let $\Phi$ be a \mrc depth four circuit of size at most $s\leq n^{\frac{t}{2}}$ that computes $\IMM_{n,d}$. Then with a probability of at least $1-o(1)$, over $V\leftarrow D$, $\Phi|_{V}$ is a \mrc depth four circuit of bottom support at most $t$ in $Z$ variables.
\end{lemma}
\begin{proof}
  Let $m$ be any monomial that is computed at the bottom layer of $\Phi$. It is easy to see that $V\leftarrow D$ sets variables across matrices independently and also variables across the rows (columns resp.) in each right (left resp.) matrix independently. That is, the probability of setting two variables is independent unless they are from the same row (column resp.) in a matrix from the right (left resp.) side of the block.
  Let us first compute the probability that a monomial survives if all the variables in its support come from a single matrix. A monomial $m$ survives of $Z$-support $t$ survives if and only if the random restriction keeps these variables alive. 
  \paragraph*{Matrices $Z^{(i,R,1)}$:} Let these $t_1,\cdots,t_{\tiln}$ be the distribution of variables across $n$ rows such that $t = t_1+\cdots+t_{\tiln}$.
  \begin{align*}
    \prob{V\leftarrow D}{m|_V \neq 0} \leq \prod_{i\in[\tiln]}\frac{{\tiln - t_i \choose \tiln^{\eta} - t_i}}{{\tiln \choose \tiln^{\eta}}} = \prod_{i\in[\tiln]}\inparen{\frac{\tiln^{\eta}}{\tiln}}^{t_i} = \tiln^{-(1-\eta)t}.
  \end{align*}
  \paragraph*{Matrices $Z^{(i,L,1)}$:} Let these $t_1,\cdots,t_{\tiln}$ be the distribution of variables across $n$ rows such that $t = t_1+\cdots+t_{\tiln}$.
  \begin{align*}
    \prob{V\leftarrow D}{m|_V \neq 0} \leq \prod_{i\in[\tiln]}\frac{{\tiln - t_i \choose \tiln^{\eta} - t_i}}{{\tiln \choose \tiln^{\eta}}} = \prod_{i\in[\tiln]}\inparen{\frac{\tiln^{\eta}}{\tiln}}^{t_i} = \tiln^{-(1-\eta)t}.
  \end{align*}
  \paragraph*{Matrices $Z^{(i,R,j)}$ for all $j\in[2, k'-\varepsilon'\log{n}]$:} Let these $t_1,\cdots,t_{\tiln}$ be the distribution of variables across $n$ rows such that $t = t_1+\cdots+t_{\tiln}$.
  \begin{align*}
    \prob{V\leftarrow D}{m|_V \neq 0} \leq \prod_{i\in[\tiln]}\inparen{\frac{2}{\tiln}}^{t_i} = \inparen{\frac{2}{\tiln}}^{t}.
  \end{align*}
  \paragraph*{Matrices $Z^{(i,L,j)}$ for all $j\in[2, k'-\varepsilon'\log{n}]$:} Let these $t_1,\cdots,t_{\tiln}$ be the distribution of variables across $n$ columns such that $t = t_1+\cdots+t_{\tiln}$.
  \begin{align*}
    \prob{V\leftarrow D}{m|_V \neq 0} \leq \prod_{i\in[\tiln]}\inparen{\frac{2}{\tiln}}^{t_i} = \inparen{\frac{2}{\tiln}}^{t}.
  \end{align*}
  \paragraph*{Matrix $Z^{(i,R,j)}$ for all $j > k'-\varepsilon'\log{n}$:} Let these $t_1,\cdots,t_{\tiln}$ be the distribution of variables across $n$ rows such that $t = t_1+\cdots+t_{\tiln}$.
  \begin{align*}
    \prob{V\leftarrow D}{m|_V \neq 0} \leq \prod_{i\in[\tiln]}\inparen{\frac{1}{\tiln}}^{t_i} = \inparen{\frac{1}{\tiln}}^{t}.
  \end{align*}
  \paragraph*{Matrix $Z^{(i,L,j)}$ for all $j > k'-\varepsilon'\log{n}$:} Let these $t_1,\cdots,t_{\tiln}$ be the distribution of variables across $n$ columns such that $t = t_1+\cdots+t_{\tiln}$.
  \begin{align*}
    \prob{V\leftarrow D}{m|_V \neq 0} \leq \prod_{i\in[\tiln]}\inparen{\frac{1}{\tiln}}^{t_i} = \inparen{\frac{1}{\tiln}}^{t}.
  \end{align*}
  Thus, we can now infer that any monomial of support $t$ stays alive with a probability of at most $p$ where
  \begin{align*}
    p = \max\inbrace{\tiln^{-(1-\eta)t},\tiln^{-t}, \inparen{\frac{2}{\tiln}}^{t}}.
  \end{align*}
  By taking a union bound, we get that with a probability of at least $1-s\cdot p = 1 - o(1)$, $\Phi|_V$ is a circuit of bottom $Z$-support at most $t$.
\end{proof}

We recall the following lemma from \cite{ChiSTACS20}. We provide the proof of this lemma in \cref{sec:upperbdckt} for the sake of completeness.
\begin{lemma}[Section 3.1, \cite{ChiSTACS20}]
  \label{lem:cktub}
  Let $N,r',m$ and $t$ be positive integers such that $m +r't < \frac{N}{2}$. Let $\Psi$ be a processed syntactic \mrc depth four circuit of size $s$ and every monomial computed at the bottom product gate has $Z$-support of at most $t$. Then, $\CM{r',m}(\Psi)$ is at most $s\cdot{\frac{2N\delta}{t}+1\choose r'}\cdot{N\choose m +r't}\cdot(m +r't)$.
\end{lemma}

\subsection*{Effect of Restrictions on $\IMM_{n,d}$}

Let $g^{(i,L)}_{1,a}$ be the $(1,a)$th entry in product of matrices $\prod_{j = 0}^{k'}(X^{(i,L,k'+1-j)})|_V$. Let $g^{(i,R)}_{b,1}$ be the $(b,1)$th entry in product of matrices $\prod_{j = 1}^{k'+1}(X^{(i,R,j)})|_V$. Let $g^{(i)}$ the $(1,1)$th entry in the product of all the matrices in the block $B_i$. Then we can express $g^{(i)}$ as follows.
\begin{align*}
  g^{(i)} = \sum_{a,b\in[n]}g^{(i,L)}_{1,a}\cdot y^{(i)}_{a,b}\cdot g^{(i,R)}_{b,1}.
\end{align*}
Let $P|_V$ obtained by restricting $\IMM_{n,d}$ with the restriction $V \leftarrow D$. Thus,
\begin{align*}
  P|_V = \prod_{i=1}^{r'}g^{(i)}\,.
\end{align*}

Thus, we can say that $g^{(i,L)}_{1,a}$ is the $(a,1)$th entry in the product $\inparen{\prod_{j = 0}^{k'}X^{(i,L,k'+1-j)}|_V}^T = \prod_{j = 1}^{k'+1}\inparen{X^{(i,L,j)}|_V}^T$. Putting this together with the structure of random restrictions, we get the following observation.
\begin{observation}
  Structures of polynomials $g^{(i,L)}_{1,a}$ and $g^{(i,R)}_{a,1}$ are similar.
\end{observation}

Let $\alpha = y^{(1)}_{(a_1,a_2)}\cdot y^{(2)}_{(a_3,a_4)}\cdot\ldots\cdot y^{(r')}_{(a_{2r'-1},a_{2r'})}$. Then,
\begin{align*}
  \partial_\alpha (P|_V) = \prod_{i=1}^{r'}g^{(i,L)}_{1,a_{2i-1}}\cdot g^{(i,R)}_{a_{2i},1}.
\end{align*}
Henceforth, without loss of generality, we will now consider that our polynomial $\partial_{\alpha}(P|_V)$ is composed of $2r'$ many $Z$-blocks, with suitable renaming.
\begin{align*}
  \partial_\alpha (P|_V) = \prod_{i=1}^{2r'}h_i\,.
\end{align*}
Note that the polynomial $\partial_\alpha(P|_V)$ is a polynomial in $\F[Z]$. From the construction, each $g_{1,a}^{(i,L)}$ or $g_{b,1}^{(i,R)}$ (where $i\in[r']$, $a,b\in[n]$) has $(n^{\eta}\cdot 2^{k'-\varepsilon'\log{n}-1})$ many monomials each. Thus, the number of monomials to in $\partial_\alpha (P|_V)$ is equal to $(n^{\eta}\cdot 2^{k'-\varepsilon'\log{n}-1})^{2r'}$ for any $\alpha \in \mathcal{M}$.

\section{Lower Bound Against Multi-$\delta$-ic Depth Four Circuit}
Let $P|_V$ be the polynomial obtained by restricting $\IMM_{n,d}$ with the restriction $V \leftarrow D$.  

\begin{observation}
  It is important to note that $\frac{\partial^{=r'}\inparen{P|_V}}{y^{(1)}_{(a_1,a_2)}\cdot y^{(2)}_{(a_3,a_4)}\cdot\ldots\cdot y^{(r')}_{(a_{2r'-1},a_{2r'})}} = \prod_{i\in[r']}g^{(i,L)}_{1,a_{2i-1}}\cdot g^{(i,R)}_{a_{2i},1}$, for any choice of $\veca\in [n]^{2r'}$ is a multilinear polynomial over just the $Z$ variables.
\end{observation}
Let $\Delta_0 \leq  2r'-r$ where $r=2\tau r'$. From \cref{clm:hammingbnd}, we get that there is a set $\mathcal{P}_{\Delta_0}\subseteq [n]^{2r'}$ of size $\frac{n^{2r'-0.5\Delta_0}}{2^{O(r')}}$ such that for any pair of vectors $\alpha, \alpha'\in \mathcal{P}_{\Delta_0}$, $\abs{\inbrace{i\mid \alpha_i = \alpha'_i}}$ is at most $r$.

Let the carefully chosen set of monomials $\M$ over $Y$ variables be defined as follows.
\begin{align*}
  \M = \inbrace{\prod_{i\in [r']}y^{(i)}_{(a_{2i-1},a_{2i})}:\alpha = (a_1,a_2, \ldots, a_{2r'})\in \mathcal{P}_{\Delta_0}}.
\end{align*}
From the definition, we can infer the following.
\begin{observation}  
  For any $\alpha\neq \alpha'\in\M$, $\partial_{\alpha}^{=r'}(P|_V)$ and $\partial_{\alpha'}^{=r'}(P|_V)$ are distinct polynomials.
\end{observation}

  For all $\alpha\in\M$, let $M(\alpha)$ be the set of monomials in the support of the polynomial $\sigma_Y(\partial_\alpha(P|_V))$. From the construction, it is easy to see that the cardinality of the set $M(\alpha)$ is the same for all $\alpha\in\M$. Let $L_2$ denote the cardinality of $M(\alpha)$ and thus $L_2 = (n^{\eta}\cdot 2^{k'-\varepsilon'\log{n}-1})^{2r'}$.

We shall henceforth use $L_1$ to denote the cardinality of $\M$ and from afore mentioned discussion and \cref{clm:hammingbnd} we get that $L_1\geq \frac{n^{(1+\tau)r'}}{2^{O(r')}}$.

With respect to the polynomial $P|_V$, let $T_1(P|_V), T_2(P|_V)$ and $T_3(P|_V)$ be as defined in \cref{sec:LeadingMonomials}. The following lemma is the crux of all our arguments. Proof of this lemma can be found in \cref{sec:calculations}.
\begin{lemma}\label{lem:calculations}
  For all $\alpha,\alpha'\in\M$ such that $\alpha\neq\alpha'$,
  \begin{enumerate}
  \item $T_1(P|_V, \alpha) = L_2\cdot {N-k\choose m}$.
  \item $\avg{V\leftarrow D}{T_2(P|_V, \alpha)} \leq L_2\cdot {N-k\choose m}\cdot 2^{O(r')}$.
  \item
    \begin{enumerate}
    \item $\avg{V\leftarrow D}{T_3(P|_V, \alpha, \alpha)}\leq L_2\cdot{N-k\choose m}\cdot 2^{O(r')}$
    \item $\avg{V\leftarrow D}{T_3(P|_V, \alpha, \alpha')} \leq L_2\cdot {N-k\choose m}\cdot 2^{O(r')}\cdot\inparen{\inparen{\frac{m}{N-m}}^{k'}+\frac{1}{n^{(1-\tau)\nu}}}^{(2r'-r)}$.
    \end{enumerate}
  \end{enumerate}
\end{lemma}

The following lemma follows from a straightforward application of Markov inequality.
\begin{lemma}\label{lem:markovT2T3}
  \begin{align*}
    \prob{V\leftarrow D}{\inparen{T_2(P|_V) < 20\cdot \avg{V'\leftarrow D}{T_2(P|_{V'})}}\wedge\inparen{T_3(P|_{V}) < 20\cdot \avg{V'\leftarrow D}{T_3(P|_{V'})}}}\geq 0.9.
  \end{align*}
\end{lemma}

\begin{lemma}[Lemma 8.8, \cite{KS14}]
  \label{lem:T2IEbound}
  With a probability of at least 0.9, there is a set $\M'\subseteq \M$ of size at least $4L_1/5$ such that for all $\alpha\in \M'$, $T_2(P|_V,\alpha) \leq 100\cdot\avg{V\leftarrow D}{T_2(P|_V, \alpha)}/L_1$.
\end{lemma}

From \cref{lem:IE-KumarSaraf}, \cref{lem:calculations} and \cref{lem:T2IEbound}, we can infer the following.
\begin{align*}
  \sum_{\substack{\alpha\in\M'\\\beta \in \M(\alpha)}}\abs{S_m(\alpha,\beta)} &= \sum_{\alpha\in\M'}T_1(P|_V,\alpha) \geq \frac{4L_1}{5}\cdot L_2\cdot {N-k\choose m}\\
  \sum_{\substack{\alpha\in\M'\\\beta\neq\gamma \in \M(\alpha)}}\abs{S_m(\alpha,\beta)\cap S_m(\alpha,\gamma)} &= \sum_{\alpha\in\M'}T_2(P|_V,\alpha) \leq  \frac{4L_1}{5}\cdot L_2\cdot {N-k\choose m}\cdot 2^{O(r')}\\
  \text{and thus}~\abs{\bigcup_{\substack{\alpha\in\M \\ \beta \in \M(\alpha)}}S_m(\alpha,\beta)} &\geq \abs{\bigcup_{\substack{\alpha\in\M' \\ \beta \in \M(\alpha)}}S_m(\alpha,\beta)} \geq \frac{L_1\cdot L_2\cdot {N-k\choose m}}{2^{O(r')}}.
\end{align*}
The last line of the math-block follows from \cref{lem:IE-KumarSaraf}.

\begin{lemma} From the setting our parameters, we get that $\inparen{\frac{m}{N-m}}^{k'}\geq n^{-\nu(1-\tau)}$, and thus
  \begin{align*}
    \avg{V\leftarrow D}{T_3(P|_V)} \leq L_1^2\cdot L_2\cdot{N-k\choose m}\cdot 2^{O(r')}\cdot\inparen{\frac{m}{N-m}}^{k'(2r'-r)}.
  \end{align*}
\end{lemma}
\begin{proof}
  By simplifying the expression $\inparen{\frac{m}{N-m}}^{k'}\geq n^{-\nu(1-\tau)}$, we get that
  \begin{align*}
    \inparen{\frac{m}{N-m}}^{k'} = \inparen{\frac{1-\Gamma}{1+\Gamma}}^{k'} \approx \frac{e^{-\Gamma\cdot k'}}{2n^\varepsilon} = \frac{e^{-\frac{\ln{n}}{\lambda\cdot k'}\cdot k'}}{2n^\varepsilon}= \frac{1}{2n^{\varepsilon+\frac{1}{\lambda}}}\approx \frac{1}{4n^{2\varepsilon}} = \frac{1}{4n^{0.5800}}
  \end{align*}
  and
  \begin{align*}
    n^{-\nu(1-\tau)} \approx n^{-0.5805}.
  \end{align*}
  The rest of the proof follows from Item 3 of \cref{lem:calculations} and simple application of linearity of expectation.
  \begin{align*}
    &\avg{V\leftarrow D}{T_3(P|_V)}\\
    &\leq \avg{V\leftarrow D}{\sum_{\alpha,\alpha'\in\M}T_3(P|_V,\alpha,\alpha')}\\
    &= \avg{V\leftarrow D}{\sum_{\alpha\in\M}T_3(P|_V,\alpha,\alpha)}+\avg{V\leftarrow D}{\sum_{\substack{\alpha,\alpha'\in\M\\\alpha\neq\alpha'}}T_3(P|_V,\alpha,\alpha')}\\
    &\leq\sum_{\alpha\in\M}L_2\cdot{N-k\choose m}\cdot 2^{O(r')}+\sum_{\substack{\alpha,\alpha'\in\M\\\alpha\neq\alpha'}}L_2\cdot {N-k\choose m}\cdot 2^{O(r')}\cdot\inparen{\inparen{\frac{m}{N-m}}^{k'}+\frac{1}{n^{(1-\tau)\nu}}}^{(2r'-r)}\\
    &\leq L_1\cdot L_2\cdot{N-k\choose m}\cdot 2^{O(r')}+ L_1^2\cdot L_2\cdot {N-k\choose m}\cdot 2^{O(r')}\cdot\inparen{\inparen{\frac{m}{N-m}}^{k'}+\frac{1}{n^{(1-\tau)\nu}}}^{(2r'-r)}\\
    &\leq L_1^2\cdot L_2\cdot{N-k\choose m}\cdot 2^{O(r')}\cdot\inparen{\inparen{\frac{m}{N-m}}^{k'}+\frac{1}{n^{(1-\tau)\nu}}}^{(2r'-r)}\\
    &\leq L_1^2\cdot L_2\cdot{N-k\choose m}\cdot 2^{O(r')}\cdot\inparen{2\inparen{\frac{m}{N-m}}^{k'}}^{(2r'-r)}\\
    &= L_1^2\cdot L_2\cdot{N-k\choose m}\cdot 2^{O(r')}\cdot\inparen{\frac{m}{N-m}}^{k'(2r'-r)}.
  \end{align*}
\end{proof}
By the setting of parameters, $L_1 = n^{(1+\tau)r'}\cdot 2^{-O(r')}$. By the choice of parameters, we get that $L_1$ is greater than $\inparen{\frac{N-m}{m}}^{k'(2r'-r)}$ and we use this to obtain the inequality in line 6 of the math-block above. With a probability of at least $0.9$, we get that
\begin{align*}
    T_3(P|_V) 
  \leq 20\cdot \avg{V'\leftarrow D}{T_3(P|_{V'})}
  \leq L_1^2\cdot L_2\cdot{N-k\choose m}\cdot 2^{O(r')}\cdot\inparen{\frac{m}{N-m}}^{k'(2r'-r)}
\end{align*}
and
\begin{align*}
  \sum_{\substack{\alpha, \alpha'\in \M'\\\beta\in M(\alpha)\\\gamma\in M(\alpha')\\(\alpha, \beta) \neq (\alpha, \gamma)}}\abs{A_m(\alpha,\beta)\cap A_m(\alpha',\gamma)} \leq \sum_{\substack{\alpha, \alpha'\in \M\\\beta\in M(\alpha)\\\gamma\in M(\alpha')\\(\alpha, \beta) \neq (\alpha, \gamma)}}\abs{A_m(\alpha,\beta)\cap A_m(\alpha',\gamma)} = T_3(P|_V)
\end{align*}
From the afore mentioned discussion and by using the Strong Inclusion Exclusion lemma (\cref{lem:IE-KumarSaraf}), we get the following.
\begin{align*}
 \CM{r',m}(P|_V) \geq \abs{\bigcup_{\substack{\alpha \in \M'\\\beta\in M(\alpha)}}A_m(\alpha,\beta)}\geq\frac{L_2\cdot {N-k\choose m}}{2^{O(r')}\cdot\inparen{\frac{m}{N-m}}^{k'(2r'-r)}}.
\end{align*}

\begin{theorem}\label{thm:lowbottomsupport}
  Let $n$, $d$ and $\delta$ be large enough integers such that $d\leq n^{0.01}$ and $\delta\leq n^{0.016}$. The polynomial $P|_V$ of degree $d= (2k'+3)\cdot r'$, over $n^{O(1)}$ variables $Y\sqcup Z$ such that any syntactically \mrc depth four circuit with bottom $Z$-support of at most $t=2\times 10^{-5}k'$, computing it must have size $n^{\Omega\inparen{r'}}$.
\end{theorem}
\begin{proof} Let $\Phi|_V$ be the \mrc depth four circuit of bottom $Z$ support of at most $t$, that computes the polynomial $P|_V$. Thus,
  \begin{align*}
    \CM{r',m}(\Phi|_V) &= \CM{r',m}(P|_V).
  \end{align*}
  Combining the above discussion with \cref{lem:cktub}, we get that
  \begin{align*}
    s\cdot{\frac{2N\delta}{t}+1\choose r'}\cdot{N\choose m +r't}\cdot(m +r't) \geq \frac{L_2\cdot {N-k\choose m}}{2^{O(r')}\cdot\inparen{\frac{m}{N-m}}^{k'(2r'-r)}}.
  \end{align*}
  Thus,
  \begin{align*}
    s &\geq \frac{L_2\cdot {N-k\choose m}}{2^{O(r')}\cdot\inparen{\frac{m}{N-m}}^{k'(2r'-r)}\cdot{\frac{2N\delta}{t}+1\choose r'}\cdot{N\choose m +r't}\cdot(m +r't)}\\
      &= \frac{L_2}{2^{O(r')}\cdot\inparen{\frac{m}{N-m}}^{k'(2r'-r)}\cdot{\frac{2N\delta}{t}+1\choose r'}}\cdot\frac{{N-k\choose m}}{{N\choose m +r't}}\\
      &\approx \frac{L_2}{2^{O(r')}\cdot\inparen{\frac{m}{N-m}}^{k'(2r'-r)}\cdot{\frac{2N\delta}{t}+1\choose r'}}\cdot\inparen{\frac{N-m}{N}}^k\cdot\inparen{\frac{m}{N-m}}^{r't} & \text{using \cref{lem:bin-gkks}}\\
      &= \frac{1}{2^{O(r')}\cdot{\frac{2N\delta}{t}+1\choose r'}}\cdot\inparen{\frac{N-m}{m}}^{k'(2r'-r)-r't} &\text{since $L_2 = \inparen{\frac{N}{N-m}}^{k}$}\\
      &\geq \frac{1}{2^{O(r')}}\cdot\inparen{\frac{N-m}{m}}^{k'(2r'-r)-r't}\cdot\inparen{\frac{tr'}{2eN\delta+t}}^{r'} &\text{since ${n\choose k}\leq \inparen{\frac{en}{k}}^{k}$}\\
      &= \frac{1}{2^{O(r')}}\cdot\inparen{\frac{N-m}{m}}^{k'(2r'-r)-r't}\cdot\inparen{\frac{t}{n^{(1+\eta)}\delta}}^{r'} &\text{since $N=O(n^{(1+\eta)}\cdot r')$}\\
      &= \frac{1}{2^{O(r')}}\cdot \inparen{\frac{1+\Gamma}{1-\Gamma}}^{k'(2r'-r)-r't}\cdot\inparen{\frac{t}{n^{(1+\eta)}\delta}}^{r'}\\
      &\geq \frac{1}{2^{O(r')}}\cdot \inparen{1+\Gamma}^{2(k'(2r'-r)-r't)}\cdot\inparen{\frac{t}{n^{(1+\eta)}\delta}}^{r'} &\text{Since $\frac{1}{1-\Gamma}\geq (1+\Gamma)$}\\
      &= \frac{1}{2^{O(r')}}\cdot \inparen{n^{\varepsilon}}^{4(1-\tau-10^{-5})r'}\cdot\inparen{\frac{t}{n^{(1+\eta)}\delta}}^{r'}& \text{since $(1+\Gamma)^{k'} = 2n^{\varepsilon'}$}\\
      &= \frac{1}{2^{O(r')}}\cdot {n}^{(4\varepsilon(1-\tau-10^{-5})-(1+\eta))r'}\cdot\inparen{\frac{t}{\delta}}^{r'}\\
    &\geq \frac{1}{2^{O(r')}}\cdot{n}^{0.017r'}\cdot{\inparen{\frac{t}{\delta}}^{r'}}.
  \end{align*}
  Thus, for all $\delta\leq n^{0.016}$, we get that $s\geq n^{\Omega(r')}$. 
\end{proof}

\subsection*{Putting it all together}

Let $\Phi$ be a \mrc depth four circuit of size at most $n^{\frac{t}{2}}$ computing the $\IMM_{n,d}$ polynomial. From \cref{lem:cktrandrest}, we get that with a probability of at least $(1-o(1))$ over $V\leftarrow D$, $\Phi|_V$ is a \mrc depth four circuit of bottom support at most $t$. Note that $\Phi|_V$ is of size at most $n^{\frac{t}{2}}$. From \cref{thm:lowbottomsupport}, $\Phi|_V$ must have size at least $n^{\Omega(r')} = n^{c_0r'}$. From our choice of parameters, $c_0r'$ is at most $\frac{t}{2}$. Thus, any \mrc depth four circuit computing $\IMM_{n,d}$ must have size at least $n^{\Omega(\sqrt{d})}$. We can now formally state our main theorem as follows.

\begin{theorem}\label{thm:maintheorem}
  Let $n$, $d$ and $\delta$ be integers such that $d\leq n^{0.01}$ and $\delta\leq n^{0.016}$. Then any syntactically \mrc depth four circuit computing $\IMM_{n,d}$ must be of size $n^{\Omega(\sqrt{d})}$.
\end{theorem}

\section{Proof of \cref{lem:calculations}}\label{sec:calculations}
\paragraph*{Proof of Item~1 of \cref{lem:calculations}:} 
Recall that that all monomials in the set $M(\alpha)$ are of degree $k=d-r'$ and $T_1(P|_V, \alpha) = \sum_{\beta \in M(\alpha)}|S_m(\alpha,\beta)|$. $S_m(\alpha,\beta)$ corresponds to the set of multilinear monomials $\gamma$ of degree $m$ whose support has zero intersection with the support of the monomial $\beta$. Thus, $\abs{S_m(\alpha, \beta)} = {N-k \choose m}$. Thus,
\begin{align*}
  T_1(P|_V, \alpha) = \sum_{\beta\in M(\alpha)}{N-k \choose m} = L_2\cdot {N-k\choose m}.
\end{align*}

\paragraph*{Proof of Item~2 of \cref{lem:calculations}:} Recall that $T_2(P|_V, \alpha)$ corresponds to the sum of cardinalities of sets of monomials $S_m(\alpha, \beta)\cap S_m(\alpha, \gamma)$, for all $\beta,\gamma \in M(\alpha)$. Further for all $\beta, \gamma\in M(\alpha)$, $S_m(\alpha, \beta)\cap S_m(\alpha, \gamma)$ corresponds to set of multilinear monomials of degree $m$ which are disjoint from both $\beta$ and $\gamma$. At most $k+\Delta(\beta,\gamma)$ variables are contained in $\beta$ and $\gamma$, and thus we get the following.
\begin{align*}
  T_2(P|_V, \alpha) &= \sum_{\substack{\beta, \gamma \in M(\alpha)\\ \beta\neq\gamma}}\abs{S_m(\alpha, \beta)\cap S_m(\alpha, \gamma)}\\
		    &= \sum_{\substack{\beta, \gamma \in M(\alpha)\\ \beta\neq\gamma}} {N-k-\Delta(\beta,\gamma)\choose m}\\
		    &= \sum_{\substack{\beta, \gamma \in M(\alpha)\\ \beta\neq\gamma}} {N-k-\Delta(\beta,\gamma)\choose m} \frac{{N-k\choose m}}{{N-k\choose m}}\\
		    &= {N-k\choose m}\sum_{\substack{\beta, \gamma \in M(\alpha)\\ \beta\neq\gamma}} \frac{{N-k-\Delta(\beta,\gamma)\choose m}}{{N-k\choose m}}\\
		    &\approx {N-k\choose m}\sum_{\substack{\beta, \gamma \in M(\alpha)\\ \beta\neq\gamma}}\inparen{\frac{N-m}{N}}^{\Delta(\beta,\gamma)} & \text{using \cref{lem:bin-gkks}}.
\end{align*}
The quantity $\Delta(\beta,\gamma)$ could be as small as $1$ and thus the quantity $\inparen{\frac{N-m}{N}}^{\Delta(\beta,\gamma)}$ could be as large as $\frac{N-m}{N}$. 
We now need the following proposition from \cite{KS14} to show that on average $\sum_{\substack{\beta, \gamma \in M(\alpha)\\ \beta\neq\gamma}}\inparen{\frac{N-m}{N}}^{\Delta(\beta,\gamma)}$ cannot be too large. Even though the proof of \cref{prop:T2Dbound} follows from Proposition 9.1 in \cite{KS14}, we present it again in \cref{app:T2Dbound} for the sake of completeness and show that it is resilient under the change of parameters.
\begin{proposition}[Proposition 9.1, \cite{KS14}] For any monomial $\beta$ in $M(\alpha)$ (where $M(\alpha)$ is the set of monomials in the support of  $\sigma_Y(\partial_\alpha(P|_V))$),
  \label{prop:T2Dbound}
  \begin{align*}
    \avg{V\leftarrow D}{\sum_{\substack{\gamma \in M(\alpha)\\ \beta\neq\gamma}}\inparen{\frac{N-m}{N}}^{\Delta(\beta,\gamma)}} \leq 2^{O(r')}.
  \end{align*}
\end{proposition}
Using \cref{prop:T2Dbound}, we get that
\begin{align*}
  \avg{V\leftarrow D}{T_2(P|_V, \alpha)} &\approx \avg{V\leftarrow D}{{N-k\choose m}\sum_{\substack{\beta, \gamma \in M(\alpha)\\ \beta\neq\gamma}}\inparen{\frac{N-m}{N}}^{\Delta(\beta,\gamma)}}\\
					 &= {N-k\choose m}\avg{V\leftarrow D}{\sum_{\beta\in M(\alpha)}\sum_{\substack{\gamma \in M(\alpha)\\ \beta\neq\gamma}}\inparen{\frac{N-m}{N}}^{\Delta(\beta,\gamma)}}\\
					 &= {N-k\choose m}\sum_{\beta\in M(\alpha)}\avg{V\leftarrow D}{\sum_{\substack{\gamma \in M(\alpha)\\ \beta\neq\gamma}}\inparen{\frac{N-m}{N}}^{\Delta(\beta,\gamma)}}\\
					 &\leq {N-k\choose m}\sum_{\beta\in M(\alpha)}\inparen{2^{O(r')}}\\
  &\leq {N-k\choose m}\cdot L_2\cdot 2^{O(r')}.
\end{align*}

\paragraph*{Proof of Item 3 of \cref{lem:calculations}:} Recall that the term $T_3(P|_V, \alpha, \alpha')$ corresponds to the sum of cardinalities of sets of monomials $A_m(\alpha,\beta)\cap A_m(\alpha',\gamma)$, for all $\beta\in M(\alpha)$ and $\gamma\in M(\alpha')$. Note that $\abs{A_m(\alpha,\beta)\cap A_m(\alpha',\gamma)}$ is upper bounded by the number of multilinear monomials of degree $m+k$ which are divisible by both $\beta$ and $\gamma$. A multilinear monomial $\gamma'$ is divisible by $\beta$ and $\gamma$ if there exists monomials $\gamma_1$ and $\gamma_2$ such that $\gamma' = \gamma_1\cdot \beta = \gamma_2\cdot \gamma$ and it must be the case that $\gamma_2$ contains all elements of $\beta\setminus\gamma$ and $\gamma_1$ contains all elements of $\gamma\setminus\beta$. The rest of the $m-\Delta(\beta,\gamma)$ many elements are the same in $\gamma_1$ and $\gamma_2$. Thus,
\begin{align*}
  T_3(P|_V, \alpha, \alpha') &= \sum_{\substack{\beta\in M(\alpha)\\\gamma\in M(\alpha')\\(\alpha, \beta) \neq (\alpha', \gamma)}}\abs{A_m(\alpha,\beta)\cap A_m(\alpha',\gamma)}\\
			     &\leq \sum_{\substack{\beta\in M(\alpha)\\\gamma\in M(\alpha')\\(\alpha, \beta) \neq (\alpha', \gamma)}}{N - k - \Delta(\beta, \gamma) \choose m - \Delta(\beta, \gamma)}\\
			     &= \sum_{\substack{\beta\in M(\alpha)\\\gamma\in M(\alpha')\\(\alpha, \beta) \neq (\alpha', \gamma)}}{N - k - \Delta(\beta, \gamma) \choose m - \Delta(\beta, \gamma)}\frac{{N-k\choose m}}{{N-k\choose m}}\\
			     &= {N-k\choose m}\sum_{\substack{\beta\in M(\alpha)\\\gamma\in M(\alpha')\\(\alpha, \beta) \neq (\alpha', \gamma)}}\frac{{N - k - \Delta(\beta, \gamma) \choose m - \Delta(\beta, \gamma)}}{{N-k\choose m}}\\
			     &= {N-k\choose m}\sum_{\substack{\beta\in M(\alpha)\\\gamma\in M(\alpha')\\(\alpha, \beta) \neq (\alpha', \gamma)}}\frac{(N-k-\Delta(\beta,\gamma))!\cdot m!}{(m-\Delta(\beta,\gamma))!\cdot (N-k)!}\\
			     &\approx {N-k\choose m}\sum_{\substack{\beta\in M(\alpha)\\\gamma\in M(\alpha')\\(\alpha, \beta) \neq (\alpha', \gamma)}}\frac{N^k\cdot m^{\Delta(\beta, \gamma)}}{N^{k+\Delta(\beta,\gamma)}} & \text{using \cref{lem:bin-gkks}}\\
			     &= {N-k\choose m}\sum_{\substack{\beta\in M(\alpha)\\\gamma\in M(\alpha')\\(\alpha, \beta) \neq (\alpha', \gamma)}}\inparen{\frac{m}{N}}^{\Delta(\beta, \gamma)}.
\end{align*}
This above sum can be further refined into two cases as follows.
\begin{align*}
  T_3^{=}(P|_V, \alpha) &= {N-k\choose m}\sum_{\substack{\beta, \gamma\in M(\alpha)\\ \beta \neq \gamma}}\inparen{\frac{m}{N}}^{\Delta(\beta, \gamma)}\\
\text{and $\forall~\alpha \neq \alpha'$, }~  T_3^{\neq}(P|_V, \alpha, \alpha') &= {N-k\choose m}\sum_{\substack{\beta\in M(\alpha)\\\gamma\in M(\alpha')\\(\alpha, \beta) \neq (\alpha', \gamma)}}\inparen{\frac{m}{N}}^{\Delta(\beta, \gamma)}.
\end{align*}

As in the case of $T_2(P|_V, \alpha)$, we can compute the expected value of $T_3^{=}(P|_V, \alpha)$ using \cref{prop:T2Dbound} as follows.
\begin{align*}
  \avg{V\leftarrow D}{T_3(P|_V, \alpha)} &\approx \avg{V\leftarrow D}{{N-k\choose m}\sum_{\substack{\beta, \gamma \in M(\alpha)\\ \beta\neq\gamma}}\inparen{\frac{m}{N}}^{\Delta(\beta,\gamma)}}\\
  &\leq \avg{V\leftarrow D}{{N-k\choose m}\sum_{\substack{\beta, \gamma \in M(\alpha)\\ \beta\neq\gamma}}\inparen{\frac{N-m}{N}}^{\Delta(\beta,\gamma)}}  &~\text{(Since $m<\frac{N}{2}$)}\\
					 &\leq {N-k\choose m}\sum_{\beta\in M(\alpha)}\inparen{2^{O(r')}} &\text{using \cref{prop:T2Dbound}}\\
  &\leq {N-k\choose m}\cdot L_2\cdot 2^{O(r')}.
\end{align*}

For any $\beta \in M(\alpha)$, let $T_3^{\neq}(P|_V, \alpha, \alpha', \beta)$ be defined as follows. 
\begin{align*}
  T_3^{\neq}(P|_V, \alpha, \alpha', \beta) = {N-k\choose m}\sum_{\substack{\gamma\in M(\alpha')\\\beta \neq  \gamma}}\inparen{\frac{m}{N}}^{\Delta(\beta, \gamma)}
\end{align*}
and thus
\begin{align*}
  T_3^{\neq}(P|_V, \alpha, \alpha') = \sum_{\beta\in M(\alpha)}T_3^{\neq}(P|_V, \alpha, \alpha', \beta).
\end{align*}
For any $\beta \in M(\alpha), \gamma\in M(\alpha')$, let $\bldiff(\beta,\gamma)$ be defined as follows.
\begin{align*}
  \bldiff(\beta,\gamma) = \inbrace{i\mid \Delta(\beta^{(i)}, \gamma^{(i)}) = k'}.
\end{align*}
For any $\beta \in M(\alpha)$, let $C_V^{t}(\beta)$ be the set of monomials $\gamma \in M(\alpha')$ such that $(\alpha, \beta) \neq (\alpha', \gamma)$ and $\abs{\bldiff(\beta,\gamma)}=t$. Note that for all $\gamma\in C_V^{t}(\beta)$, $\Delta(\beta, \gamma)\geq t\cdot k'$. 
\begin{align*}
  T_{3}^{\neq}(P|_V, \alpha, \alpha', \beta) &= {N-k\choose m}\sum_{\substack{\gamma\in M(\alpha')\\(\alpha, \beta) \neq (\alpha', \gamma)}}\inparen{\frac{m}{N}}^{\Delta(\beta, \gamma)}\\
							  &= {N-k\choose m}\sum_{\substack{\gamma\in M(\alpha')}}\inparen{\frac{m}{N-m}}^{\Delta(\beta, \gamma)}\cdot\inparen{\frac{N-m}{N}}^{\Delta(\beta, \gamma)}\\
							  &\leq {N-k\choose m}\sum_{t=0}^{2r'-r}\sum_{\substack{\gamma\in C_V^{t}(\beta)}}\inparen{\frac{m}{N-m}}^{t\cdot k'}\cdot\inparen{\frac{N-m}{N}}^{\Delta(\beta, \gamma)}\\
  &\leq {N-k\choose m}\sum_{t=0}^{2r'-r}\inparen{\frac{m}{N-m}}^{t\cdot k'}\cdot\inparen{\sum_{\substack{\gamma\in C_V^{t}(\beta)}}\inparen{\frac{N-m}{N}}^{\Delta(\beta, \gamma)}}\\
\end{align*}
First inequality follows from the fact that $\frac{m}{N-m}< 1$ and $\Delta(\beta, \gamma) \geq t\cdot k'$.

Recall that $\Delta(\beta,\gamma) = \sum_{i\in [2r']}\Delta(\beta^{(i)},\gamma^{(i)})$. Let $C_V^t(\beta)$ be expressed as a sum of all $C_V^{t,S}(\beta)$ for all sets $S$ of size $2r'-r-t$ that indexes blocks $i$ such that $\alpha_i \neq \alpha'_i$ and $\bldiff(\beta^{(i)}, \gamma^{(i)})<k'$.

\begin{align*}
  &\sum_{\substack{\gamma\in C_V^{t}(\beta)}}\inparen{\frac{N-m}{N}}^{\Delta(\beta, \gamma)}\\
  &= \sum_{\substack{\gamma\in C_V^{t}(\beta)}}\prod_{i\in[r']}\inparen{\frac{N-m}{N}}^{\Delta(\beta^{(i)}, \gamma^{(i)})}\\
  &= \sum_{\substack{S\subseteq [2r'-r]\\\abs{S}=2r'-r-t}}\sum_{\substack{\gamma\in C_V^{t,S}(\beta)}}\prod_{i\in[r']}\inparen{\frac{N-m}{N}}^{\Delta(\beta^{(i)}, \gamma^{(i)})}\\
  &\leq {2r'-r\choose t}\cdot\max_{\substack{S\subseteq [2r'-r]\\\abs{S}=2r'-r-t}}\inbrace{\inparen{\prod_{i\in S}\sum_{\substack{\gamma\in C_V^{t,S}(\beta)}}\inparen{\frac{N-m}{N}}^{\Delta(\beta^{(i)}, \gamma^{(i)})}}\cdot\inparen{\prod_{i\notin S}\sum_{\substack{\gamma\in C_V^{t,S}(\beta)}}\inparen{\frac{N-m}{N}}^{\Delta(\beta^{(i)}, \gamma^{(i)})}}}.
\end{align*}

From the definition, it follows that for all those $\gamma\in C_V^t(\beta)$ and $i\in S$ (for any $S$), we trivially get that $\gamma^{(i)}\in A_V^{(i)}(\beta)$ where $A_V^{(i)}(\beta^{(i)})$ be the set of $\gamma^{(i)}$ such that there is some $j\in[k'-1]$ such that $\beta^{(i,j)}=\gamma^{(i,j)}$ and $\beta^{(i,j+1)}=\gamma^{(i,j+1)}$ (see \cref{app:T2Dbound}). 
\begin{align*}
  \avg{V\leftarrow D}{\prod_{i\in S}\sum_{\substack{\gamma\in C_V^{t,S}(\beta)}}\inparen{\frac{N-m}{N}}^{\Delta(\beta^{(i)}, \gamma^{(i)})}}&=\prod_{i\in S}\avg{V\leftarrow D}{\sum_{\substack{\gamma\in C_V^{t,S}(\beta)}}\inparen{\frac{N-m}{N}}^{\Delta(\beta^{(i)}, \gamma^{(i)})}}\\
  &\leq \inparen{\frac{1}{n^\nu}}^{2r'-r-t}.
\end{align*}
For the blocks that are not indexed by $S$, we get the following crude bound.
\begin{align*}
  \avg{V\leftarrow D}{\prod_{i\notin S}\sum_{\substack{\gamma\in C_V^{t,S}(\beta)}}\inparen{\frac{N-m}{N}}^{\Delta(\beta^{(i)}, \gamma^{(i)})}}&\leq 2^{O(r')} & \text{using \cref{prop:T2Dbound}}.
\end{align*}

Putting these together we get that
\begin{align*}
  \avg{V\leftarrow D}{\sum_{\substack{\gamma\in C_V^{t}(\beta)}}\inparen{\frac{N-m}{N}}^{\Delta(\beta, \gamma)}}\leq{2r'-r\choose t}\cdot \frac{2^{O(r')}}{n^{(2r'-r-t)\nu}}
\end{align*}
and
\begin{align*}
  &\avg{V\leftarrow D}{T_{3}^{\neq}(P|_V, \alpha, \alpha', \beta)}\\
  &= \avg{V\leftarrow D}{{N-k\choose m}\sum_{t=0}^{2r'-r}\inparen{\frac{m}{N-m}}^{t\cdot k'}\cdot\inparen{\sum_{\substack{\gamma\in C_V^{t}(\beta)}}\inparen{\frac{N-m}{N}}^{\Delta(\beta, \gamma)}}}\\
  &= {N-k\choose m}\cdot\sum_{t=0}^{2r'-r}\inparen{\frac{m}{N-m}}^{t\cdot k'}\cdot\avg{V\leftarrow D}{\sum_{\substack{\gamma\in C_V^{t}(\beta)}}\inparen{\frac{N-m}{N}}^{\Delta(\beta, \gamma)}}\\
  &\leq {N-k\choose m}\cdot\sum_{t=0}^{2r'-r}\inparen{\frac{m}{N-m}}^{t\cdot k'}\cdot{2r'-r\choose t}\cdot \frac{2^{O(r')}}{n^{(2r'-r-t)\nu}}\\
  &= {N-k\choose m}\cdot 2^{O(r')}\cdot \sum_{t=0}^{2r'-r}{2r'-r\choose t}\cdot\inparen{\frac{m}{N-m}}^{t\cdot k'}\cdot \frac{1}{n^{(2r'-r-t)\nu}}\\
  &= {N-k\choose m}\cdot 2^{O(r')}\cdot\inparen{\inparen{\frac{m}{N-m}}^{k'}+\frac{1}{n^{\nu}}}^{2r'-r}.
\end{align*}
Thus,
\begin{align*}
  \avg{V\leftarrow D}{T_{3}^{\neq}(P|_V, \alpha, \alpha')} &= \sum_{\beta\in M(\alpha)}\avg{V\leftarrow D}{T_{3}^{\neq}(P|_V, \alpha, \alpha', \beta)}\\
  &\leq L_2\cdot{N-k\choose m}\cdot 2^{O(r')}\cdot\inparen{\inparen{\frac{m}{N-m}}^{k'}+\frac{1}{n^{\nu}}}^{2r'-r}.
\end{align*}

\section{Proof of \cref{prop:T2Dbound}}\label{app:T2Dbound}

Recall that each $\beta\in M(\alpha)$ can be expressed as a product of $\beta^{(i)}$'s for all $i\in[2r']$. That is, for all $i\in[2r']$, $\beta^{(i)}$ corresponds to a contribution to \emph{monomial} $\beta$ from the $i$th block. It is easy to see that $\Delta(\beta, \gamma)$ is equal to $\sum_{i\in[2r']}\Delta(\beta^{(i)}, \gamma^{(i)})$ as the blocks are defined over disjoint sets of variables. Thus,
\begin{align*}
  \avg{V\leftarrow D}{\sum_{\substack{\gamma \in M(\alpha)\\ \beta\neq\gamma}}D^{-\Delta(\beta,\gamma)}} &\leq \prod_{i\in[r']}\avg{V\leftarrow D}{\sum_{\substack{\gamma \in M(\alpha)\\ \beta\neq\gamma}}D^{-\Delta(\beta^{(i)},\gamma^{(i)})}}.
\end{align*}
To prove \cref{prop:T2Dbound}, it is sufficient to show that each term of the right side of the above equation is at most $5$.
\begin{lemma}\label{lem:AVbetai-app}
  For all $\beta\in M(\alpha)$ and all $i\in[2r']$,
  \begin{align*}
    \avg{V\leftarrow D}{\sum_{\substack{\gamma \in M(\alpha)\\ \beta\neq\gamma}}D^{-\Delta(\beta^{(i)},\gamma^{(i)})}} \leq O(1).
  \end{align*}
\end{lemma}
\begin{proof}
  Let $M^{(i)}(\alpha)$ be the set of monomials corresponding to the polynomial $h_i$ after deriving $P|_V$ with $\alpha\in\M$.
  \begin{align*}
    \partial_\alpha (P|_V) = \prod_{i\in[2r']}h_i\,.
  \end{align*}
  It is now clear that $M(\alpha) \subseteq \prod_{i=1}^{2r'}M^{(i)}(\alpha)$. For all $\beta^{(i)} \in M^{(i)}(\alpha)$, let 
  \begin{itemize}
  \item $A_V^{(i)}(\beta^{(i)})$ be the set of $\gamma^{(i)}$ such that there is some $j\in[k'-1]$ such that $\beta^{(i,j)}=\gamma^{(i,j)}$ and $\beta^{(i,j+1)}=\gamma^{(i,j+1)}$.
    \item $B_V^{(i)}(\beta^{(i)})$ be the set of $\gamma^{(i)}$ such that there is some $j,j'\in[k']$ such that if $\beta^{(i,j)}=\gamma^{(i,j)}$ and $\beta^{(i,j')}\neq\gamma^{(i,j')}$ then $j<j'$.
  \end{itemize}
  \begin{claim}\label{clm:BVbetai}
    For all $\beta^{(i)}\in M^{(i)}(\alpha)$,
    \begin{align*}
      \avg{V\leftarrow D}{\sum_{\gamma^{(i)} \in B_V^{(i)}(\beta^{(i)})}D^{-\Delta(\beta^{(i)}, \gamma^{(i)})}} \leq O(1).
    \end{align*}
  \end{claim}
  \begin{proof}
    Let $B_V^{(i)}(\beta^{(i)})$ be further partitioned into $k'+1$ sets $B_V^{(i,j)}(\beta^{(i)})$ for all integers $t\in[0,k']$ such that $B_V^{(i,t)}(\beta^{(i)})$ consists of all those $\gamma^{(i)}$'s such that $\beta^{(i,j)}=\gamma^{(i,j)}$ for all $j\leq t$ and $\beta^{(i,j)}\neq\gamma^{(i,j)}$ for all $j>t$. This means that every $\gamma^{(i)}$ in $B_V^{(i,t)}(\beta^{(i)})$ matches with $\beta^{(i)}$ at the first $t$ positions and differ at every position after that. Thus, it is easy to see that there are at most $\prod_{j>t}\deg_V(X^{(i,j)})$ such $\gamma_i$'s. This is due to the fact that there are $\deg_V(X^{(i,j)})$ choices at every position $j>t$. Recall that $\prod_{j\in[k']}\deg_V(X^{(i,j)}) = D^{k'}$ and for all $\gamma^{(i)} \in B_V^{(i,t)}(\beta^{(i)})$, $\Delta(\beta^{(i)}, \gamma^{(i)}) = k'-t$. Thus,
    \begin{align*}
      \sum_{\gamma^{(i)} \in B_V^{(i,t)}(\beta^{(i)})}D^{-\Delta(\beta^{(i)}, \gamma^{(i)})} &\leq \prod_{j>t}\deg_V(X^{(i,j)})\cdot D^{-(k'-t)}\\
    \end{align*}
    For $t=0$, the above expression equals $1$. For $t=1$, the above expression is at most $\frac{D}{ n ^{\eta}}$. For $t\in [2, k'-\varepsilon'\log{n}]$, the above expression evaluates to
    \begin{align*}
      \sum_{\gamma^{(i)} \in B_V^{(i,t)}(\beta^{(i)})}D^{-\Delta(\beta^{(i)}, \gamma^{(i)})} &\leq D^{t}\prod_{j>t}(\deg_V(X^{(i,j)}))^{-1}\\
											     &= \frac{D}{ n ^{\eta}}\prod_{j=2}^t\frac{D}{2}\\
      &\leq \frac{D}{ n ^{\eta}} &\text{Since $D<2$}.
    \end{align*}
    For $t>k'-2\log{n}$, $\deg_V(X^{i,j})=1$ for all $j>t$ and thus swe get that
    \begin{align*}
      \sum_{\gamma^{(i)} \in B_V^{(i,t)}(\beta^{(i)})}D^{-\Delta(\beta^{(i)}, \gamma^{(i)})} \leq \prod_{j>t}\deg_V(X^{(i,j)})\cdot D^{-(k'-t)} = D^{-(k'-t)}.
    \end{align*}
    Putting it all together, we get
    \begin{align*}
      \sum_{\gamma^{(i)} \in B_V^{(i)}(\beta^{(i)})}D^{-\Delta(\beta^{(i)}, \gamma^{(i)})} &= \sum_{t=0}^{k'}\sum_{\gamma^{(i)} \in B_V^{(i,t)}(\beta^{(i)})}D^{-\Delta(\beta^{(i)}, \gamma^{(i)})}\\
											   &\leq 1 + (k'-\varepsilon'\log{n})\cdot\frac{D}{ n ^{\eta}}+\sum_{t = k'-\varepsilon'\log{n}+1}^{k'}D^{-(k'-t)}\\
											   &\leq 2+\sum_{t=0}^{\varepsilon'\log{n}}D^{-t}\\
      &\leq 2+\frac{D}{D-1} \leq O(1).
    \end{align*}
    The second inequality in the above math-block uses the fact that $2(k'-\varepsilon'\log{n}) <  n ^\eta$. This completes the proof of \cref{clm:BVbetai}.
  \end{proof}
  \begin{claim}\label{clm:AVbetai}
    For all $\beta^{(i)}\in M^{(i)}(\alpha)$,
    \begin{align*}
      \avg{V\leftarrow D}{\sum_{\gamma^{(i)} \in A_V^{(i)}(\beta^{(i)})}D^{-\Delta(\beta^{(i)}, \gamma^{(i)})}} \leq \frac{1}{n^\nu}.
    \end{align*}
  \end{claim}
  \begin{proof}
    For any $\beta^{(i)}\in M^{(i)}(\alpha)$ and $\gamma^{(i)}$, and for some $j\in[k']$ we call $j$ an agreement switch if    $\beta^{(i,j-1)}\neq\gamma^{(i,j-1)}$ and $\beta^{(i,j)}=\gamma^{(i,j)}$, and a disagreement switch if $\beta^{(i,j-1)}=\gamma^{(i,j-1)}$ and $\beta^{(i,j)}\neq\gamma^{(i,j)}$.
    
    Let $A_V^{(i)}(\beta^{(i)})$ be further partitioned into $k'$ sets $A_V^{(i,j)}(\beta^{(i)})$ for all $t\in[0,k']$ based on the number of switches such that $A_V^{(i,t)}(\beta^{(i)})$ contains all those $\gamma^{(i)}$ that contain $t$ many disagreements with respect to $\beta^{(i)}$.

    Let $S_t$ be a $t$ sized subset of $[k']$ and $b\in\{0,1\}$ be a bit. By specifying the first switch $b$ and the switch locations $S_t$, we can precisely characterize any element in $A_V^{(i,t)}(\beta^{(i)})$. Let $T = \inbrace{d_1, d_2, \cdots, d_s}$ be the set of $s$ coordinates where $\gamma^{(i)}\in A_V^{(i,t)}(\beta^{(i)})$ disagrees with $\beta^{(i)}$ where $\gamma^{(i)}$ is characterized by $(b,S_t)$. 
    Note that there are at most $\deg_V(X^{(i,d_j)})$ many options for picking an edge to the next layer, that is, at most $\deg_V(X^{(i,d_j)})$ many possible labels for a position $j\in[s]$.

    Further, if there is a agreement switch, then the last disagreeing edge label must have the same end point as that of $\beta^{(i)}$. Since the edge end points are picked uniformly at random in $D$ and $\beta^{(i)}$ being fixed, the probability that the correct endpoint is chosen is $\frac{1}{ n }$.
    
    Thus given a disagreement pattern $T$ and the fact that there are $q$ agreement switches, the number of $\gamma^{(i)}$'s that match the disagreement pattern and $q$ agreement switches is at most $\prod_{j\in T}\deg_V(X^{(i,j})\cdot n ^{-q}$. If there are $t$ switches in a $\gamma^{(i)}$ the number of agreement switches can at most be $(t+1)/2$ and at least be $\max\inbrace{1,(t-1)/2}$. This is due to the fact that the agreement and disagreement switches alternate. Thus the number of elements in $A_V^{(i,t)}(\beta^{(i)})$ with disagreement pattern $T$ (denoted by $A_V^{(i,t,T)}(\beta^{(i)})$) is at most $\prod_{j\in T}\deg_V(X^{(i,j})\cdot n ^{-\max\inbrace{1,(t-1)/2}}$. Recall that fixing $(b,S_t)$ fixes an element of $A_V^{(i,t)}(\beta^{(i)})$ and thus it fixes a disagreement pattern. There are at most $2\cdot {k'\choose t}$ many sets $(b,S_t)$ and thus there are at most $2\cdot {k'\choose t}$ disagreement patterns. Putting this all together we get that
    \begin{align*}
      &\avg{V\leftarrow D}{\sum_{\gamma^{(i)} \in A_V^{(i)}(\beta^{(i)})}D^{-\Delta(\beta^{(i)}, \gamma^{(i)})}}\\
      &\leq \sum_{\substack{t\in[k']\\T\subseteq[k']\\\gamma^{(i)}\in A_V^{(i,t,T)}(\beta^{(i)})}}\inparen{\prod_{j\in T}\deg_V(X^{(i,j})}\cdot n ^{-\max\inbrace{1,(t-1)/2}}\cdot D^{-\abs{T}}\,.
    \end{align*}
    \begin{claim}
      For all $i\in[r']$ and all $T\subseteq[k']$, $\inparen{\prod_{j\in T}\deg_V(X^{(i,j})}\cdot D^{-\abs{T}} \leq n^{\varepsilon'}$.
    \end{claim}
    \begin{proof}
      It is easy to see that the product $\inparen{\prod_{j\in T}\deg_V(X^{(i,j)})}$ is maximized for $T=[k'-\varepsilon'\log{n}]$.
      \begin{align*}
	\max_T\inbrace{\inparen{\prod_{j\in T}\deg_V(X^{(i,j)})}\cdot D^{-\abs{T}}}&\leq \inparen{\prod_{j=1}^{k'-\varepsilon'\log{n}}\deg_V(X^{(i,j)})}\cdot D^{\varepsilon'\log{n}-k'}\\
										  &= D^{k'}\cdot D^{\varepsilon'\log{n}-k'}\\
	&= D^{\varepsilon'\log{n}}\leq n^{\varepsilon'}.
      \end{align*}
    \end{proof}
    Using this fact, we get that
    \begin{align*}
      &\avg{V\leftarrow D}{\sum_{\gamma^{(i)} \in A_V^{(i)}(\beta^{(i)})}D^{-\Delta(\beta^{(i)}, \gamma^{(i)})}}\\ &\leq \sum_{\substack{t\in[k']}}2\cdot {k'\choose t}\cdot \frac{n^{\varepsilon'}}{ n ^{\max\inbrace{1,(t-1)/2}}}\\
      &= \frac{2\cdot k'\cdot n^{\varepsilon'}}{n}+\frac{ k'(k'-1)\cdot n^{\varepsilon'}}{n}+\frac{ k'(k'-1)(k-2)\cdot n^{\varepsilon'}}{3n} + \sum_{t\geq 4}{k'\choose t}\cdot\frac{2\cdot n^{\varepsilon'}}{ n ^{(t-1)/2}}\\
	&= O\inparen{\frac{k'^3\cdot n^{\varepsilon'}}{n}} + \sum_{t\geq 4}{k'\choose t}\cdot\frac{2\cdot n^{\varepsilon'}}{ n ^{(t-1)/2}}
    \end{align*}
    For the setting of values of $k' = \Theta(\sqrt{d})$ where $d\leq n^{0.01}$, $\varepsilon'=0.34$ and $\nu=0.631$, we can verify that for all $t\geq 4$,
    \begin{align*}
      {k'\choose t}\cdot\frac{2\cdot n^{\varepsilon'}}{ n ^{(t-1)/2}} \leq O\inparen{\frac{k'^3\cdot n^{\varepsilon'}}{n}}\,.
    \end{align*}
    Thus, we get the following bound.
    \begin{align*}
      \avg{V\leftarrow D}{\sum_{\gamma^{(i)} \in A_V^{(i)}(\beta^{(i)})}D^{-\Delta(\beta^{(i)}, \gamma^{(i)})}} \leq  O\inparen{\frac{k'^4\cdot n^{\varepsilon'}}{n}}\leq \frac{1}{n^\nu}\,.
    \end{align*}
    This is the only inequality that forces the degree parameter $d$ to have a value of at most $n^{0.01}$.
  \end{proof}
  By putting \cref{clm:BVbetai} and \cref{clm:AVbetai}, we get the needed result.
  \begin{align*}
    &\avg{V\leftarrow D}{\sum_{\substack{\gamma \in \M(\alpha)\\ \beta\neq\gamma}}D^{-\Delta(\beta^{(i)},\gamma^{(i)})}}\\
    &= \avg{V\leftarrow D}{\sum_{\substack{\gamma\in A_V^{(i)}(\beta^{(i)})}}D^{-\Delta(\beta^{(i)},\gamma^{(i)})}} + \avg{V\leftarrow D}{\sum_{\substack{\gamma \in B_V^{(i)}(\beta^{(i)})}}D^{-\Delta(\beta^{(i)},\gamma^{(i)})}}\\
    &\leq \frac{1}{n^\nu} + O(1) \leq O(1).
  \end{align*}
\end{proof}

\section*{Acknowledgement}
Research supported by CHE-PBC Post Doctoral Fellowship, and Binational Science Foundation Grant of mentor Noga Ron-Zewi. We would like to thank Nutan Limaye and Srikanth Srinivasan for patiently listening to a preliminary presentation of \cite{ChiSTACS20} (upon which this work is based), Srikanth Srinivasan for his suggestions that eventually led to the current version of this work, Mrinal Kumar for helping us understand their work~\cite{KS14} better, and Ramprasad Saptharishi for lucidly presenting the crux of the arguments in \cite{KS14} in his survey~\cite{github}.
\bibliography{ref}
\appendix

\section{Upper Bound on $\CM{r',m}(\Psi)$}\label{sec:upperbdckt}
Recall that $\Psi$ is a sum of at most $s$ many products of polynomials $T^{(1)}+\cdots+T^{(s)}$ where each $T_i$ is a syntactically \mrc product of polynomials of low monomial support.

We shall first prove a bound on $\CM{r',m}(T_i)$ for an arbitrary $T_i$ and derive a bound on $\CM{r',m}(C)$ by using sub-additivity of the measure~(cf.\ \cref{prop:subadd}).

Let $T$ be a syntactic \mrc product of polynomials $\tilde{Q}_1(Y,Z)\cdot\tilde{Q}_2(Y,Z)\cdot\ldots \cdot\tilde{Q}_D(Y,Z)\cdot R(Y)$ such that $\abs{\mathrm{MonSupp}_Z\inparen{\tilde{Q}_i}} \leq t$. We will first argue that $D$ is not too large since $T$ is a syntactically \mrc product. We shall first pre-process the product $T$ by doing the following procedure.

Repeat this process until all but at most one of the factors in $T$ (except $R$) have a $Z$-support of at least $\frac{t}{2}$.
\begin{enumerate}
\item Pick two factors $\tilde{Q}_{i_1}$ and $\tilde{Q}_{i_2}$ such that they have the
  smallest $Z$-support amongst $\tilde{Q}_1,\cdots, \tilde{Q}_D$.
\item If both of them have support strictly less than $\frac{t}{2}$, merge these factors to obtain a new factor, and repeat step 1. Else, stop.
\end{enumerate}

In the afore mentioned procedure, it is important to note that the monomial support in $Z$ variables post merging will still be at most $t$ since the factors being merged are of support strictly less than $\frac{t}{2}$. Henceforth, W.L.O.G we shall consider that every product gate at the top, in any \mrc depth four circuit to be in a processed form.

Let $T = Q_1(Y,Z)\cdot Q_2(Y,Z)\cdot\ldots\cdot Q_D(Y,Z)\cdot R(Y)$ be the product obtained after the preprocessing. All but one of the $Q_i$'s have a $Z$-support of at least $\frac{t}{2}$. The total $Z$-support is at most $\abs{Z}\delta=N\delta$ since $T$ is a syntactically \mrc product. Thus $D$ could at most be $\frac{2N\delta}{t}+1$.

\begin{lemma}\label{lem:cm-cktub}
  Let $N,r', m$ and $t$ be positive integers such that $m+r't < \frac{N}{2}$. Let $T$ be a processed syntactic \mrc product of polynomials $Q_1(Y,Z)\cdot Q_2(Y,Z)\cdot\ldots\cdot Q_D(Y,Z)\cdot R(Y)$ such that $\abs{\mathrm{MonSupp}_Z\inparen{Q_i}} \leq t$. Then, $\CM{r',m}(T)$ is at most ${D\choose r'}\cdot{N\choose m+r't}\cdot(m+r't)$.
\end{lemma}

Before proving \cref{lem:cm-cktub}, we shall first use it to show an upper bound on the dimension of the space of Projected Shifted Skew Partial derivatives of a depth four \mrc circuit of low bottom support.
\begin{proof}[Proof of \cref{lem:cktub}]
  W.L.O.G we can assume that $C$ be expressed as $\sum_{i=1}^s T^{(i)}$ such that $T^{(i)}$ is a processed syntactically \mrc product of polynomials with  bottom support at most $t$ with respect to $Z$ variables. From \cref{prop:subadd}, we get that $\CM{r',m}(C) \leq \sum_{i=1}^s\CM{r',m}(T^{(i)})$. From the afore mentioned discussion we know that the number of factors in $T^{(i)}$ with a non-zero $Z$-support is at most $\frac{2N\delta}{t}$. From \cref{lem:cm-cktub}, we get that $\CM{k,\ell}(T^{(i)})$ is at most ${\frac{2N\delta}{t}+1\choose r'}\cdot{N\choose m+r't}\cdot(m+r't)$.	By putting all of this together, we get that
  \begin{align*}
    \CM{r',m}(C) \leq s\cdot{\frac{2Nr}{t}+1\choose r'}\cdot{N\choose m+r't}\cdot(m+r't)\,.
  \end{align*}
\end{proof}

\begin{proof}[Proof of \cref{lem:cm-cktub}]
  We will first show by induction on $r'$, the following.
  \begin{align*}
    \partial^{=r'}_Y T \subseteq &\fspan\inbrace{\bigcup_{S\in{[D]\choose D-r'}}\inbrace{\prod_{i\in S} Q_i(Y,Z)\cdot \vecz^{\leq r't}_{\ml}\cdot \F[Y]}} \\&\qquad\bigcup\fspan\inbrace{\bigcup_{S\in{[D]\choose D-r'}}\inbrace{\prod_{i\in S} Q_i(Y,Z)\cdot \vecz^{\leq r'\delta t}_{\nml}\cdot \F[Y]}}
  \end{align*}
  The base case of induction for $r'=0$ is trivial as $T$ is already in the required form. Let us assume the induction hypothesis for all derivatives of order $<r'$. That is, $\partial^{=r'-1}_Y T$ can be expressed as a linear combination of terms of the form 
  \begin{align*}
    h(Y,Z) = \prod_{i\in S}Q_i(Y,Z)\cdot h_1(Z)\cdot h_2(Y)
  \end{align*}
  where $S$ is a set of size $D-(r'-1)$, $h_1(Z)$ is a polynomial in $Z$ variables of degree at most $(r'-1)\delta t$, and $h_2(Y)$ is some polynomial in $Y$ variables. In fact, $h_1(Z)$ can be expressed as a linear combination of multilinear monomials of degree at most $(r'-1)t$, and non-multilinear monomials of degree at most $(r'-1)\delta t$.

  For some $u\in[\abs{Y}]$ and some fixed $i_0$ in $S$, 
  \begin{align*}
    \frac{\partial h(Y,Z)}{\partial y_u} =&\inparen{\sum_{j\in S}\prod_{\substack{i\in S\\i\neq j}}Q_i(Y,Z)\cdot \frac{\partial Q_j(Y,Z)}{\partial y_u}\cdot h_1(Z)\cdot h_2(Y)} \\
					  &\qquad\qquad+ \frac{\prod_{i\in S}Q_i}{Q_{i_0}}\cdot Q_{i_0}(Y,Z)\cdot h_1(Z)\cdot\frac{\partial h_2(Y)}{\partial y_k}\\
    \in & \fspan\inbrace{\prod_{\substack{i\in S\\i\neq j}}Q_i(Y,Z)\cdot \frac{\partial Q_j(Y,Z)}{\partial y_u}\cdot h_1(Z)\cdot \F[Y] \mid j\in[S]}\\ & \qquad\qquad \bigcup\fspan\inbrace{\frac{\prod_{i\in S}Q_i}{Q_{i_0}}\cdot Q_{i_0}(Y,Z)\cdot h_1(Z)\cdot \F[Y]}\\
    \subseteq & \fspan\inbrace{\bigcup_{T\in{S\choose \abs{S}-1}}\inbrace{\prod_{i\in T}Q_i(Y,Z)\cdot \vecz^{\leq t}_{\ml}\cdot h_1(Z)\cdot \F[Y]}}\\ & \qquad\qquad\bigcup \inbrace{ \bigcup_{T\in{S \choose \abs{S}-1}}\inbrace{\prod_{i\in T}Q_i(Y,Z)\cdot \vecz^{\leq t\delta}_{\nml}\cdot h_1(Z)\cdot \F[Y]}}\\
    \subseteq& \fspan\inbrace{\bigcup_{T\in{[D]\choose D-r'}}\inbrace{\prod_{i\in T} Q_i(Y,Z)\cdot \vecz^{\leq r't}_{\ml}\cdot \F[Y]}} \\& \qquad\qquad\bigcup\fspan\inbrace{\bigcup_{T\in{[D]\choose D-r'}}\inbrace{\prod_{i\in T} Q_i(Y,Z)\cdot \vecz^{\leq r'\delta t}_{\nml}\cdot \F[Y]}}
  \end{align*}

  The last inclusion follows from the fact that $h_1(Z)$ is a linear combination of multilinear monomials of degree at most $(r'-1)t$, and non-multilinear monomials of degree at most $(r'-1)\delta t$. 
  From the discussion above we know that any polynomial in $\partial^{=r'}_Y(T)$ can be expressed as a linear combination of polynomials of the form $\frac{\partial h}{\partial y_u}$. Further every polynomial of the form $\frac{\partial h}{\partial y_u}$ belongs to the set 
  \begin{align*}
    W &= \fspan\inbrace{\bigcup_{T\in{[D]\choose D-r'}}\inbrace{\prod_{i\in T} Q_i(Y,Z)\cdot \vecz^{\leq r't}_{\ml}\cdot \F[Y]}}\\ &\qquad\qquad\bigcup\fspan\inbrace{\bigcup_{T\in{[D]\choose D-r'}}\inbrace{\prod_{i\in T} Q_i(Y,Z)\cdot \vecz^{\leq r'\delta t}_{\nml}\cdot \F[Y]}}.
  \end{align*}
  Thus, we get that $\partial^{=r'}T$ is a subset of $W$. This completes the proof by induction.
  
  From the afore mentioned discussion, we can now derive the following expressions.
  \begin{align*}
    \sigma_Y\inparen{\partial^{=r'}_Y T} \subseteq &\fspan\inbrace{\bigcup_{S\in {[D]\choose D-r'}}\inbrace{\prod_{\substack{i\in S}}\sigma_Y(Q_i)\cdot \vecz^{\leq r't}_{\ml}}}\\ &\qquad\qquad\bigcup \fspan\inbrace{\bigcup_{S\in {[D]\choose D-r'}}\inbrace{\prod_{\substack{i\in S}}\sigma_Y(Q_i)\cdot \vecz^{\leq r'\delta t}_{\nml}}}
  \end{align*}
  \begin{align*}
    \vecz^{= m}\cdot\sigma_Y\inparen{\partial^{=r'}_Y T} \subseteq &\fspan\inbrace{\bigcup_{S\in {[D]\choose D-r'}}\inbrace{\prod_{\substack{i\in S}}\sigma_Y(Q_i)\cdot \vecz^{\leq m+r't}_{\ml}}}\\ &\qquad\qquad\bigcup \fspan\inbrace{\bigcup_{S\in {[D]\choose D-r'}}\inbrace{\prod_{\substack{i\in S}}\sigma_Y(Q_i)\cdot \vecz^{\leq m+r'\delta t}_{\nml}}}
  \end{align*}
  \begin{align*}
    \implies \fspan\inbrace{\mult\inparen{\vecz^{=m}\cdot \sigma_Y\inparen{\partial^{=r'}_Y T}}}\subseteq \fspan\inbrace{\bigcup_{S\in {[D]\choose D-r'}}\inbrace{\mult\inparen{\prod_{\substack{i\in S}}\sigma_Y(Q_i)}\cdot \vecz^{\leq m+r't}_{\ml}}}.
  \end{align*}
  
  Thus we get that $\dim\inparen{\fspan\inbrace{\mult\inparen{\vecz^{= m}\cdot\sigma_Y(\partial^{=r'}_Y T)}}}$ is at most 
  \begin{align*}
    &\dim\inparen{\fspan\inbrace{\bigcup_{S\in {[D]\choose D-r'}}\inbrace{\mult\inparen{\prod_{\substack{i\in S}}\sigma_Y(Q_i)}\cdot \vecz^{\leq m+r't}_{\ml}}}}\\
    \leq & \dim\inparen{\fspan\inbrace{\bigcup_{S\in{[D]\choose D-r'}}\inbrace{\mult\inparen{\prod_{i\in S}\sigma_Y(Q_i)}}}}\cdot \dim\inparen{\fspan\inbrace{\vecz^{
	   \leq m+r't}_{\ml}}}\\
    \leq & {D\choose D-r'}\cdot \sum_{i=0}^{m+r't}{N\choose i}\\
    \leq & {D\choose r'}\cdot {N\choose m+r't}\cdot (m+r't) & \text{(Since $m+r't < N/2$)}.
  \end{align*}
\end{proof}
\section{Missing Proofs}
\begin{proof}[Proof of \cref{clm:hammingbnd}]
  There are $n^{2r'}$ elements in $\mathcal{P}$. It is easy to see that the volume of a Hamming Ball of radius $\Delta_0/2$ for vectors of length $2r'$ is at most $\sum_{i=0}^{\Delta_0}{2r' \choose i}\cdot n^{i} \leq \frac{\Delta_0}{2}{2r' \choose \Delta_0/2}n^{\Delta_0/2}$ and thus there are at most $\frac{\Delta_0}{2}{2r' \choose 0.5\Delta_0}n^{0.5\Delta_0}$ many vectors $\veca$ in that Hamming ball. Thus, there exists a packing of these Hamming balls in $\mathcal{P}$ with at least $\frac{2n^{2r'-0.5\Delta_0}}{\Delta_0{2r'\choose 0.5\Delta_0}}$ many balls. Distance between the centers of these balls is $\Delta_0$.
\end{proof}
\end{document}

%% file: minimacros.tex
\newcommand{\inparen }[1]{\left(#1\right)}             
\newcommand{\inbrace }[1]{\left\{#1\right\}}           
\newcommand{\insquare}[1]{\left[#1\right]}             

\newcommand{\abs}[1]{\left|#1\right|}                  

\newcommand{\fspan}{\operatorname{\F\text{-span}}}

\newcommand{\zo}{\inbrace{0,1}}                        

\newcommand{\F}{\mathbb{F}}

\newcommand{\prob}[2]{\operatorname{\mathbb{P}}_{#1}\insquare{#2}}

\newcommand{\IMM}{\mathsf{IMM}}

\newcommand{\veca}{\mathbf{a}}

\newcommand{\vecz}{\mathbf{z}}

\newcommand{\SPSP}{\Sigma\Pi\Sigma\Pi}

\newcommand{\SPSPsupp}[1]{\Sigma\Pi\Sigma\Pi^{\{#1\}}}

\newcommand{\VNP}{\mathsf{VNP}}

\newcommand{\LM}{\mathsf{LM}}

\renewcommand{\epsilon}{\varepsilon}

\renewcommand{\epsilon}{\varepsilon}
\newcommand{\ignore}[1]{}